\documentclass[12pt,draftclsnofoot, onecolumn]{IEEEtran}

\usepackage{graphicx}
\usepackage{epstopdf}
\usepackage[cmex10]{amsmath}
\usepackage{amsfonts}
\usepackage{amssymb}
\usepackage{hyperref}
\usepackage{bbm,dsfont}
\usepackage{array}
\usepackage{cite}
\usepackage{amsthm}
\usepackage{pifont}
\usepackage{epstopdf}
\usepackage{pifont}
\usepackage{varwidth}
\usepackage{xcolor}
\usepackage[normalem]{ulem}
\usepackage[caption=false]{subfig}

\usepackage{algorithm}
\usepackage{algorithmicx}
\usepackage{algpseudocode}

\DeclareMathOperator*{\argmin}{argmin}
\DeclareMathOperator*{\find}{find}

\begin{document}

\title{Load-Based On/Off Scheduling for Energy-Efficient Delay-Tolerant 5G Networks}


\author{Haluk \c{C}elebi, Yavuz Yap{\i}c{\i}, \.{I}smail G\"{u}ven\c{c}, and Henning Schulzrinne
\thanks{H.~\c{C}elebi is with the Department of Electrical and Computer Engineering, Columbia University, New York, NY (e-mail:~haluk@ee.columbia.edu).

Y.~Yap{\i}c{\i} and \.{I}.~G\"{u}ven\c{c} are with the Department of Electrical and Computer Engineering, North Carolina State University, Raleigh, NC (e-mail:~\{yyapici,iguvenc\}@ncsu.edu).

H.~Schulzrinne is with the Department of Computer Science, Columbia University, New York, NY (e-mail:~hgs@cs.columbia.edu).
}
\thanks{This research was supported in part by NSF under the grant CNS-1814727.}
}%

\maketitle

\begin{abstract}

Dense deployment of small cells is seen as one of the major approaches for addressing the traffic demands in next-generation 5G wireless networks. The energy efficiency, however, becomes a key concern together with this massive amount of small cells. In this study, we therefore consider the energy-efficient small cell networks (SCN) using smart on/off scheduling (OOS) strategies, where a certain fraction of small base stations (SBS) are put into less energy-consuming sleeping states to save energy. To this end, we first represent the overall SCN traffic by a new \textit{load} variable, and analyze its statistics rigorously using Gamma approximation. We then propose two novel OOS algorithms exploiting this load variable in centralized and distributed fashions. We show that proposed load based OOS algorithms can lead to as high as $50\%$ of energy savings without sacrificing the average SCN throughput. In addition, load based strategies are shown to work well under high SCN traffic and delay-intolerant circumstances, and can be implemented efficiently using the load statistics. We also show that the performance of load based algorithms gets maximized for certain length of sleeping periods, where assuming short sleep periods is as energy-inefficient as keeping SBSs in sleep states for very long.

\end{abstract}


\section{Introduction}

Massive densification of small cell networks (SCNs) is commonly seen as one of the major pillars of 5G wireless networks  to cope with the ever-increasing mobile data traffic~\cite{bhushan2014network,quek2013small}. For such dense deployments of SCNs, developing dynamic cell management and user-access mechanisms are crucial for saving energy at off-peak hours and for boosting the throughput of the  network~\cite{hwang2013holistic,samarakoon2016ultra}. Active cells not only consume energy, but also increase interference in the communication environment. Therefore, green and energy-efficient strategies that opportunistically place cells into sleep mode becomes important for  unplanned cell locations, especially with dynamically varying user distributions, spatial load, and traffic load.

Due to user mobility and varying traffic demand, number of small cells that are required to satisfy the quality of service requirements (QoS) of users change continuously. In particular, numerous types of user equipment (UEs) such as tablets, mobile phones, gaming consoles, e-readers, and machine type devices cause heterogeneous traffic patterns. This heterogeneous UE traffic environment necessitates energy-efficient and dynamic techniques where the small cell base stations (SBSs) switch to sleep mode when they are not needed, and they can also dynamically get activated when the demand is high.

While there are several recent techniques in the literature for energy-efficient small cells~\cite{samarakoon2016ultra,li2014throughput,soh2013energy,merwaday2016optimisation}, energy savings can be further improved by dynamic switching based on short-term service demand, and integrating energy-efficient sleep mode techniques with flexible access strategies for UEs. For example, in~\cite{GreenBSN:EnergyPropCell}, the geographical  area is divided into multiple grids. In each grid area, maximum number of SBSs are selected at times of peak traffic to satisfactorily serve all users. In idle periods, a subset from these selected SBSs are kept active and remaining SBSs are turned off. This strategy yields up to 53\% energy savings in dense areas,  and 23\% in sparse areas. In~\cite{samarakoon2016ultra}, reinforcement learning techniques are used for dynamically placing cells into sleep mode while simultaneously satisfying quality of service requirements.

In \cite{5GAccess:linkProvisioning}, authors study the problem of controlling multiple sets of access points each serving one UE, in a way that largest number of UEs experience  the maximum service quality. It is assumed that a UE can connect multiple cells and switch between multi-path and single path during its service period. The flexibility in terms of number of cells accessed, or accessing further cell instead of close-by cell opens a new venue for research that has not been well investigated. While delay tolerant networks (DTN) have been studied extensively in the literature in~\cite{zeng2013directional,he2013emd,cao2013routing}, to our best knowledge, it has not been explored rigorously in the context of energy-efficient SCNs, where placing certain small cells into sleep mode can save energy at the cost of latency for certain users.


In this work, which is a rigorously extended version of \cite{loadBasedOnOffHaluk}, we study energy-efficient on/off scheduling (OOS) strategies for SBSs in next-generation 5G networks. Considering a user-centric approach, where SBS selection mechanism is managed at UEs, we propose a novel \textit{load} based OOS framework with a promise of more energy-efficient SCNs. Our specific contributions are listed as follows.
\begin{itemize}
    
    \item[--] We propose a novel framework for SCNs involving randomly distributed SBSs and UEs, where the traffic load of the overall network is represented by a random \textit{load} variable. We investigate the distribution of this load variable, and derived closed form analytical expressions of respective probability distribution function (PDF) and cumulative distribution function (CDF), which are verified through extensive simulations. 
    
    \item[--] Towards achieving energy-efficient SCNs, we propose two load based (LB) OOS algorithms, where certain fraction of SBSs with relatively lower load values are put into less energy consuming (i.e., sleeping) states for a random duration of time. In particular, we introduce \textit{centralized} LB (CLB) and \textit{distributed} LB (DLB) as two novel OOS algorithms. Although CLB necessitates the knowledge of instantaneous load values of all SBSs, DLB, instead, relies on the CDF of the load requiring much fewer load values. The numerical results verify that CLB and its computational-efficient alternative DLB have very close performance.
    
    \item[--] We also consider two benchmark OOS techniques, which are \textit{random on/off} (ROO) and \textit{wake-up control} (WUC). 
    While ROO is a simple baseline algorithm~\cite{Natarajan2016SmaCel}, WUC is a more complex sophisticated algorithm requiring full-control of the macro base station (MBS) dynamically. The numerical results verify that CLB and DLB are superior to ROO, and have similar performance as WUC. 
    Furthermore, as the overall SCN traffic increases, WUC turns out to be less energy-efficient than either CLB or DLB. 
    
\end{itemize}


The rest of this paper is organized as follows. Section~\ref{sec:system} introduces the system model for SCNs with dynamic on/off operation of SBSs. Section~\ref{sec:derivations} analytically derives the traffic load distribution for a given UE using a Gamma distribution approximation. Section~\ref{sec:algorithms} proposes the centralized and distributed strategies to conduct on/off operation of SBSs. Section~\ref{sec:simulations} presents numerical results, and Section~\ref{sec:conclusion} concludes the paper.

\section{System Model}\label{sec:system}
In this section, we first overview the network model, then describe the novel load based model for the network traffic, and finally describe the power consumption model of the SBSs.

\subsection{SCN Model}\label{sec:scn_model}
We consider a densely packed SCN where low-power SBSs are operated to deliver mobile data to UEs of interest. We assume that SBSs and UEs are distributed randomly over a $2$-dimensional horizontal plane following the homogeneous Poisson point process (HPPP) with densities $\rho_{\rm c}$ and $\rho_{\rm u}$, respectively. Each UE is assumed to be able to receive service from any cell separated by at most the threshold distance $R_{\rm th}$. In addition, UEs generate traffic at random time intervals, and request to offload a file where the file size and the service request intervals have exponential distribution with rates $\lambda_{\rm F}$ and $\lambda_{\rm U}$, respectively.

\begin{figure}[!t]
	\centering
	\includegraphics[width=0.6\textwidth]{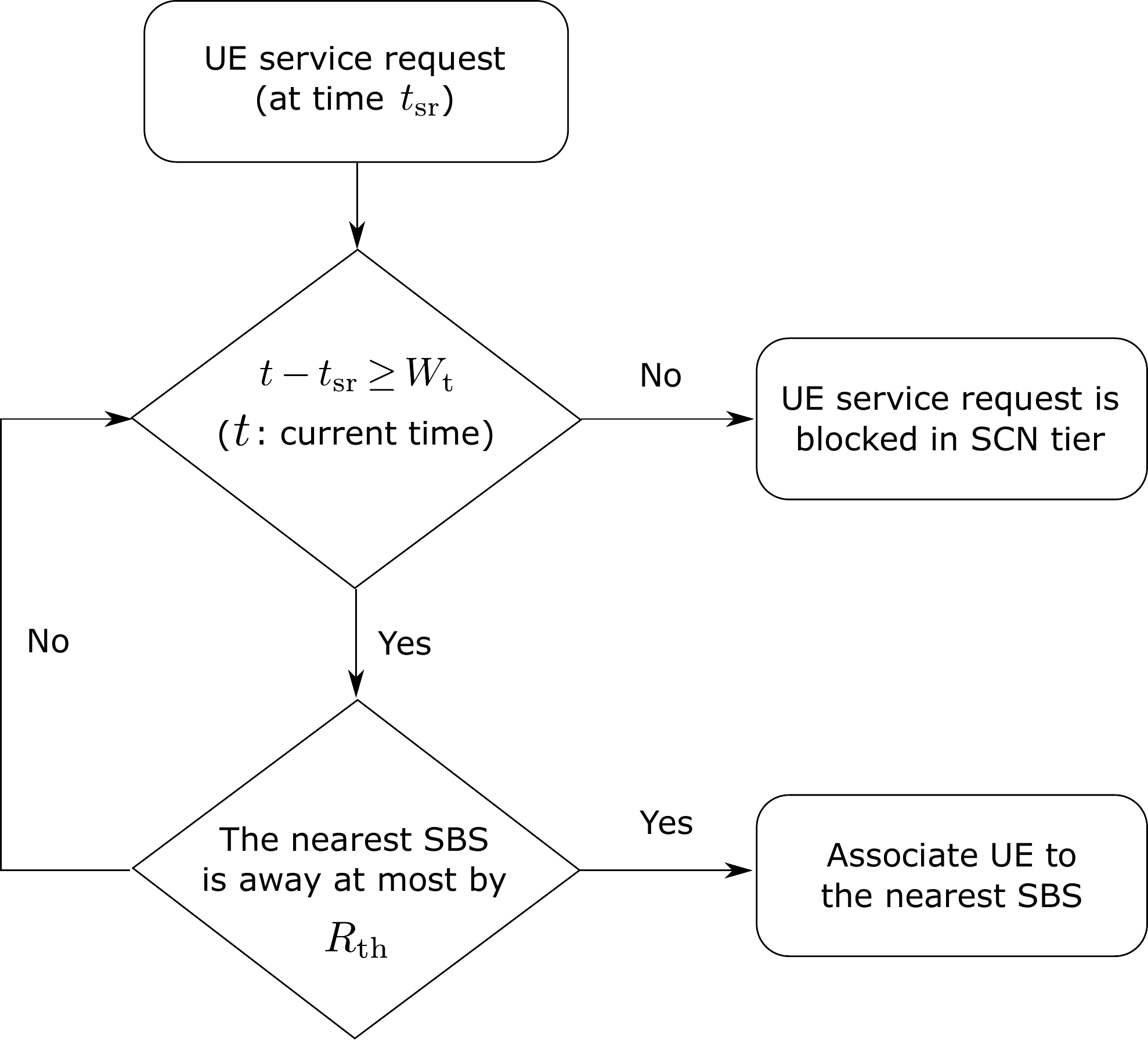}
	\caption{Delayed access strategy of UEs in SCN.}
	\label{fig:delayed_access}
\end{figure}

Considering that each UE is not involved in transmission all the time (due to exponentially distributed service request times), the energy efficiency of the overall network is desired to be improved by putting some of the SBSs into less energy-consuming (i.e., sleeping) states. Leaving details of sleeping states and the associated OOS strategies to Section~\ref{sec:algorithms}, each SBS in sleeping states is assigned with a random sleep time $T_{\rm s}$, which follows exponential distribution with rate $\lambda_{\rm S}$. The delayed access strategy under consideration is given in Fig.~\ref{fig:delayed_access}, which assumes that any UE has tolerable delay of at most $w_{\rm t}$ seconds (i.e., \textit{waiting time}). If a UE with active service request finds at least one available (i.e., \textit{idle}) SBS within the threshold distance $R_{\rm th}$ and the waiting time $w_{\rm t}$, it connects to the best (i.e., \textit{nearest}) of these SBSs to offload its desired traffic. Otherwise, it connects to MBS, and the current service request is assumed to be \textit{blocked} in SCN tier. 

In terms of interaction between UEs and SBSs, we assume that UEs do not know the location of sleeping SBSs. But rather, UEs have the perfect knowledge of distance to each non-sleeping SBSs, which can be estimated by monitoring/processing the downlink reference signals from these SBSs. The association between UEs and SBSs is set up such that each SBS serves a single UE at a time, and each UE does not change its SBS till the current service request is completely fulfilled. In addition, UEs use all available bandwidth once connected to an SBS, and quickly finish their service resulting in short service times. We finally note that SCN handles only the data traffic, and the voice traffic is handled efficiently by MBSs in macrocell tier.    

\subsection{SBS/UE Densities and Traffic Load}
We define $n_{\rm c}$ and $n_{\rm u}$ as range-dependent SBS and UE densities, respectively, which refer to average number of SBSs and UEs within a circular area of radius $R_{\rm th}$. Since location distribution for SBSs and UEs both follow HPPP, the respective Poisson distribution with the range-dependent SBS and UE densities are defined with the mean values $\nu_{\rm c} \,{=}\, \rho_{\rm c} \pi R_{\rm th}^2$ and $\nu_{\rm u} \,{=}\, \rho_{\rm u} \pi R_{\rm th}^2$, respectively. The probability that $n_{\rm c}$ SBSs and $n_{\rm u}$ UEs are present in the circular area of radius $R_{\rm th}$ are therefore given as $p_{\rm c}(i) \,{=}\, P\{n_{\rm c} \,{=}\, i\} \,{=}\, \frac{\nu_{\rm c}^i e^{-\nu_{\rm c}}}{i!}$ and  $p_{\rm u}(j) \,{=}\, P\{n_{\rm u} \,{=}\, j\} \,{=}\, \frac{\nu_{\rm u}^j e^{-\nu_{\rm u}}}{j!}$, respectively.

We define the \emph{load factor} for the $j$th UE as follows:
\begin{align}\label{eqn:load_factor_def}
w_j = \begin{cases}
	  \frac{1}{n(j)} &\mbox{if } n(j) >0,  \\
	  0 & \mbox{if } n(j) = 0,
      \end{cases}
\end{align}
where $n(j)$ is the number of SBSs that the $j$th UE can receive service (i.e., away by at most $R_{\rm th}$). Accordingly, \emph{load value} $L_i$ for the $i$th SBS is defined to be the sum of load factors associated with each UE off the $i$th SBS by at most a distance of $R_{\rm th}$, and is given as follows:
\begin{align}\label{eqn:load_def}
L_i = \sum_{j=1 }^{\infty} w_j  \mathsf{1}(i,j),
\end{align}
where  $\mathsf{1}(i,j)$  is the indicator function which is $1$ if $i$th SBS and $j$th UE are within $R_{\rm th}$ distance, and zero otherwise. 

\begin{figure}[!t]
\centering
\includegraphics[width=0.7\textwidth]{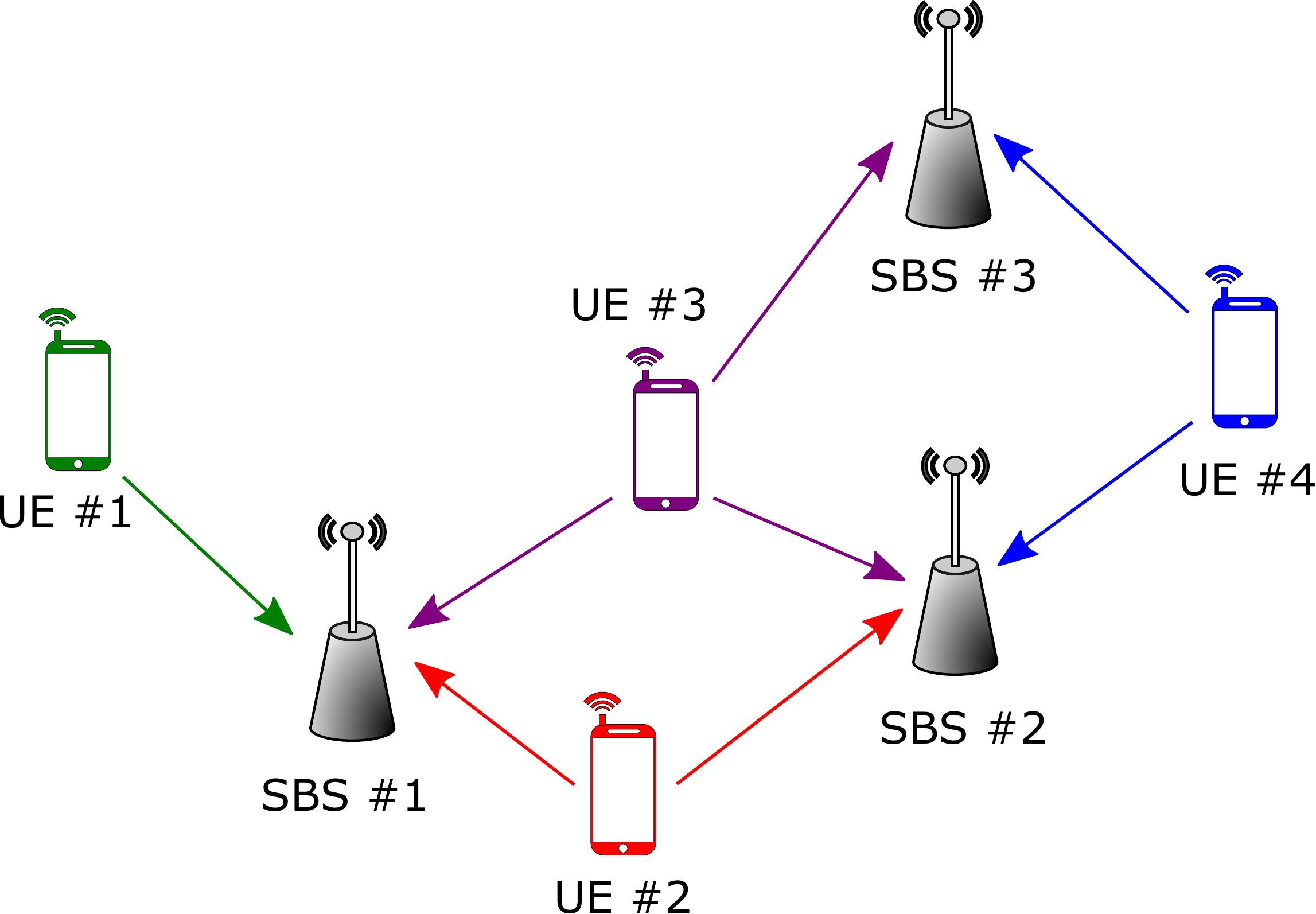}
\caption{A representative network of 3 SBSs and 4 UEs. The arrows indicate the SBSs that each UEs can receive service.}
\label{fig:load_example}
\end{figure}

As an example, we consider a representative network given in Fig.~\ref{fig:load_example}. Defining $\mathcal{S}_{i}$ as the indices of SBSs that $i$th UE can receive service, we have $\mathcal{S}_{1}\,{=}\,\{1\}$, $\mathcal{S}_{2}\,{=}\,\{1,2\}$, $\mathcal{S}_{3}\,{=}\,\{1,2,3\}$, and $\mathcal{S}_{4}\,{=}\,\{2,3\}$. Using, (\ref{eqn:load_factor_def}), load factors of UEs are computed as $w_1\,{=}\,1$, $w_2\,{=}\,\frac{1}{2}$, $w_3\,{=}\, \frac{1}{3}$, $w_4\,{=}\,\frac{1}{2}$. The respective load values of SBSs are then given using~\eqref{eqn:load_def} by $L_1\,{=}\,1\,{+}\,\frac{1}{2}\,{+}\,\frac{1}{3} \,{=}\,\frac{11}{6}$, $L_2\,{=}\,\frac{1}{2}\,{+}\,\frac{1}{3}\,{+}\,\frac{1}{2} \,{=}\,\frac{4}{3}$, and $L_3\,{=}\,\frac{1}{3}\,{+}\,\frac{1}{2}\,{=}\,\frac{5}{6}$.

We note that instantaneous load value of any SBS possibly varies with the blocked calls, UE-SBS association policies, traffic patterns, and transmission rates in a particular UE-SBS topology. Therefore, the load value of a SBS may not represent exact load distribution perfectly, but is successful enough in giving a good measure of how much traffic any specific SBS handles. Besides, distributed nature of load computation allows the design and implementation of distributed algorithms which will be discussed in following sections. 

\subsection{Power Consumption Model for SBSs}
We now briefly present the power consumption model for an arbitrary SBS, which is an important measure while evaluating the energy efficiency of the overall network. Note that since the transmission range of UEs is very limited, and respective delayed access strategy described in Fig~\ref{fig:delayed_access} is the same for all OOS strategies, average transmit power of UE is expected to be invariant in all schemes. We therefore do not include the power consumption of UEs in this study, and take into account power consumption of SBSs only.

Considering a standard BS architecture, we assume that the hardware is composed of three blocks: microprocessor (i.e., to manage radio protocols, backhaul connection, etc.), field-programmable gate array (FPGA) (i.e., to process necessary baseband algorithms), radio frequency (RF) front-end (e.g., power amplifiers, transmitter elements, etc.)~\cite{Natarajan2016SmaCel, Ashraf2011SleMod, Vereecken2012TheEff}. In order to obtain power saving (i.e., \textit{sleeping}) states, OOS strategies consider to turn off a fraction of SBSs not actively engaged in transmission. This can be done by turning off some or all of the hardware blocks, where it takes more time to boot up as more hardware blocks are turned off (i.e., deeper the state is). 

\begin{table}[ht]
\caption{SBS States, Boot-Up Times, and Power Consumption Levels}
\label{tab:sbs_states}
\centering
\begin{tabular}{cccc}
\hline
SBS State   & Boot-up time (s)  & Power Consumption (\%) \\
\hline\hline
Active 		& $0$  		& $100$ \\
Idle 		& $0$ 		& $50$	\\
Standby	    & $0.5$		& $50$ 	\\
Sleep 		& $10$		& $15$ 	\\
Off			& $30$      & $0$ 	\\
\hline
\end{tabular}
\end{table}

In Table~\ref{tab:sbs_states}, we list the SBS states considered in this work together with respective boot-up times and normalized power consumption levels, which are available in the literature~\cite{Natarajan2016SmaCel, Ashraf2011SleMod, Vereecken2012TheEff}. The description and assumptions for the SBS states are as follows.
\begin{itemize}
    \item[-] \textit{Active}: The SBS is actively engaging in transmission with full power.
    \item[-] \textit{Idle}: The SBS is ready to transmit immediately, but not transmitting currently. Hence, RF front-end is not running, and the power consumption is therefore 50\% of active state. 
    \item[-] \textit{Standby}: In this light sleep state, the heater for oscillator is turned off intentionally, and RF front-end is not running at all.
    \item[-] \textit{Sleep}: The SBS is in a deep sleep with only necessary hardware parts (power supply, central processor unit (CPU), etc.) are up.
    \item[-] \textit{Off}: The SBS is completely offline.
\end{itemize}
Note that, the sleeping state should be put into either sleep or off states to achieve significant power savings, where the respective minimum boot-up time is $10$ seconds. Since any sleeping SBS should be available right after its random sleep time $T_{\rm s}$ expires, it is not possible to put any SBS into either sleep or off states if $T_{\rm s} \,{<}\,10$ seconds. We assume that such SBSs are put into standby state, as shown in Table~\ref{tab:sleep_state_choice}, to capture the effect of turning off procedure, and meet the requirement to wake up immediately after $T_{\rm s}$ seconds. In addition, the power consumption during boot-up period is equal to that of the standby state since that particular SBS does not actively communicate with users.          

Although deeper sleeping states provide more power savings, respective longer boot-up times result in UE service requests being blocked more in SCN tier. To effectively handle this fundamental tradeoff between energy consumption and boot-up time, the optimal sleep state should be selected based on UE's delay tolerance, transmit range, and cell density. The optimal sleep state choice is beyond the scope of this study. Instead, we prefer a simple rule which puts each SBS into the deepest state as much as possible for maximum power savings  and is given below. Sleep times up to 30 seconds are stand-by or sleep which are determined by hardware limitations. However, if the sleep time is greater than 30 seconds, then, SBS can be either in sleep or off mode. Decision between sleep or off mode is made by minimum power consumption rule by taking into consideration of both the power consumption during boot-up, $p_{\text{boot-up}}$ , and sleep mode $p_{\text{sleep}}$.



\begin{table}[ht]
\caption{Sleep State Choice Based on Sleep Time ($T_{\rm s}$)}
\label{tab:sleep_state_choice}
\centering
\begin{tabular}{cc}
\hline
Sleep State   & Sleep Time ($T_{\rm s}$) (s) \\
\hline\hline
Stand-by	& $T_{\rm s} \leq 10$	\\
Sleep   	& $10 < T_{\rm s} \leq 30$	\\
Sleep 		& $T_{\rm s} >30, 10 \, p_{\text{boot-up}} \,{+}\, \left(T_{\rm s}-10\right) \, p_{\text{sleep}} < 30 \, p_{\text{boot-up}}$    \\
Off			& $T_{\rm s} >30, 10 \, p_{\text{boot-up}} \,{+}\, \left(T_{\rm s}\,{-}\,10\right) \, p_{\text{sleep}} > 30 p_{\text{boot-up}}$ \\
\hline
\end{tabular}
\end{table}


\section{Analysis of Traffic Load Distribution} \label{sec:derivations}

In this section, we analyze the distribution of the load variable as a successful measure of the actual traffic loads of SBSs. There are several studies in the literature where fitting distributions are used instead of deriving exact distributions, especially for Poisson Voronoi cell topologies~\cite{tanemura2003statistical,ferenc2007size,davies1974size}. Following a similar approach, we analyze distribution of the load variable $L$ by considering the Gamma distribution, which is verified to have satisfactory fitting performance. 

The PDF of the gamma distribution can be expressed in terms of shape parameter $\alpha$ and inverse scale parameter $\beta$ as follows:
\begin{align}
    f(x;\alpha,\beta) = \frac{\beta^\alpha x^{\alpha-1} e^{-\beta x}}{\Gamma (\alpha)},
\end{align}
where $\Gamma(\cdot)$ is the gamma function~\cite{Ross2009IntProb}. Our goal is, therefore, to determine suitable expressions  of the gamma parameters $\alpha$ and $\beta$ in terms of SCN parameters $\rho_{\rm u}$, $\rho_{\rm c}$, and $R_{\rm th}$. When the load variable $L$ is assumed to be gamma-distributed with parameters $\alpha$ and $\beta$, the first and second moments are given as 
\begin{align}
\mathbb{E}[L] &= \frac{\alpha}{\beta}, \qquad \mathbb{E}[L^2] = \frac{\alpha\left(1+\alpha \right)}{\beta^2},
\end{align}
and the parameters to be determined can be expressed as
\begin{align}
\alpha & = \beta \mathbb{E}[L], \label{eqn:gamma_alpha}\\
\beta & = \frac{\mathbb{E}[L]}{\mathbb{E}[L^2]-\mathbb{E}[L]}. \label{eqn:gamma_beta}
\end{align}
As a result, the first and the second moments of $L$ completely specifies the desired fitting distribution, and the rest of our analysis is therefore devoted to finding these moments.

\subsection{First Moment of Load Variable}
\begin{figure}[!t]
\centering
\subfloat[Single UE] {\includegraphics[width=0.35\textwidth]{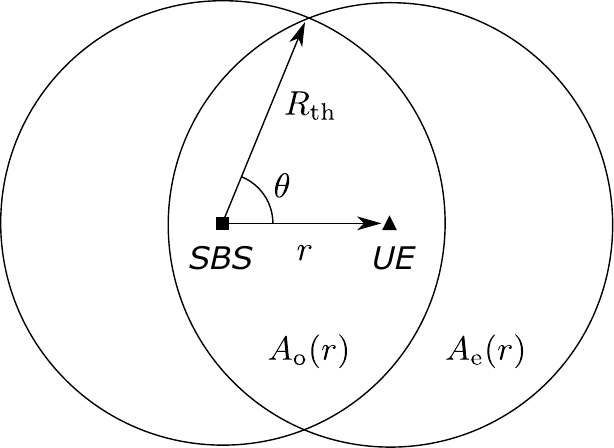}
\label{fig:oneUE}}\hspace{0.2in}
\subfloat[Two UEs] {\includegraphics[width=0.33\textwidth]{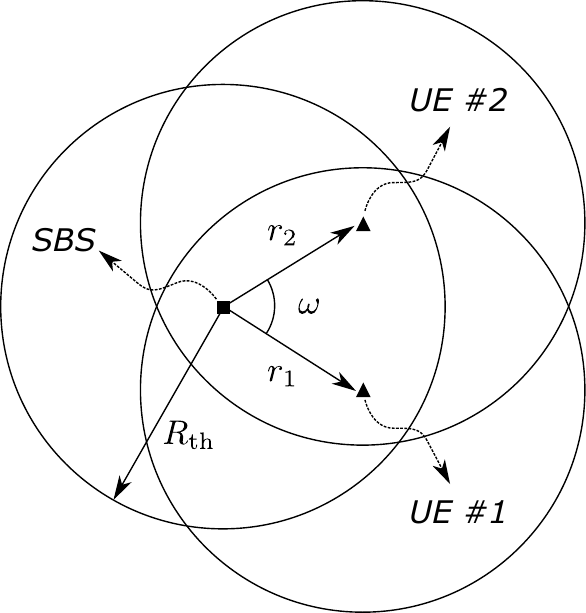}
\label{fig:twoUE}}		
\caption{A representative SCN involving a single SBS at the origin, and arbitrary UEs off by at most $R_{\rm th}$.}
\label{fig:subnetwork}
\end{figure}

The first moment of the load variable $L$ for arbitrary SBS in the network is derived by focusing on a representative sub-network shown in Fig.~\ref{fig:subnetwork}\subref{fig:oneUE}. In this framework, the target SBS (for which the load will be computed) is assumed to be located at the origin together with $n_{\rm c}$ additional SBSs and $n_{\rm u}$ UEs, which are distributed randomly over a circular area of radius $R_{\rm th}$.

The first moment of the load $L$ can be expressed as a conditional sum over all possible number of SBSs and UEs as follows:
\begin{align}\label{eqn:1st_moment_1}
	\mathbb{E}\left[ L \right] & = \sum_{i=0}^{\infty}\sum_{j=0}^{\infty} \mathbb{E} \left[ L \mid n_{\rm c} = i, n_{\rm u} = j \right] \, p_{\rm c}(i) \, p_{\rm u}(j) ,
\end{align}
and using the load definition of \eqref{eqn:load_def} in \eqref{eqn:1st_moment_1} yields
\begin{align}
\label{eqn:1st_moment_2}
	\mathbb{E}\left[ L \right] & = \sum_{i=0}^{\infty}\sum_{j=0}^{\infty} \mathbb{E} \left[ \sum_{k=1}^{j}w_{k} | n_{\rm c} = i \right] \, p_{\rm c}(i)\, p_{\rm u}(j), \\
\label{eqn:1st_moment_3}	
	& = \sum_{i=0}^{\infty}\sum_{j=0}^{\infty} \sum_{k=1}^{j} \mathbb{E} [ w_{k} | n_{\rm c} = i ]  \, p_{\rm c}(i)\, p_{\rm u}(j) .
\end{align}

We observe that the individual load factors in \eqref{eqn:1st_moment_3} (i.e., $w_k$'s) are not necessarily the same since the number of SBSs which are away from each UE by at most $R_{\rm th}$ may not be the same. The expected values of the load factors are, however, the same (i.e., $\mathbb{E} [ w_k | n_{\rm c} \,{=}\, i ] \,{=}\, \mathbb{E} [ w | n_{\rm c} \,{=}\, i ]$ for $\forall \, k$) since SBSs are distributed uniformly. We may therefore rearrange (\ref{eqn:1st_moment_3}) to obtain
\begin{align}
\label{eqn:1st_moment_4}
\mathbb{E}\left[ L \right] & = \sum_{i=0}^{\infty}\sum_{j=0}^{\infty} j \, \mathbb{E} [ w | n_{\rm c} = i ]\, p_{\rm c}(i)\, p_{\rm u}(j), \\
\label{eqn:1st_moment_5}
	& = \sum_{i=0}^{\infty} \mathbb{E} [ w | n_{\rm c} = i ] \, p_{\rm c}(i) \, \sum_{j=1}^{\infty} \, j \, p_{\rm u}(j) .
\end{align}
Realizing that the last summation in (\ref{eqn:1st_moment_5}) is the definition of the expected value for the number of user (i.e., $n_{\rm u}$), which is Poisson distributed with rate $\nu_{\rm u}$, we obtain
\begin{align}
\label{eqn:1st_moment_6}
\mathbb{E}\left[ L \right] & = \nu_{\rm u} \sum_{i=0}^{\infty} \mathbb{E} [ w | n_{\rm c} = i]\, p_{\rm c}(i),
\end{align}
which reduces to the problem of finding \textit{average traffic load} contributed by a \textit{single UE}.

In order to compute the average load factor conditioned on the number of cell (i.e., $\mathbb{E} [ w | n_{\rm c} \,{=}\, i ]$), we choose an arbitrary UE that is off the origin (i.e., the target SBS of interest) by a distance $r$ with $r \,{\leq}\,R_{\rm th}$, as shown in Fig.~\ref{fig:subnetwork}\subref{fig:oneUE}. Because any user can only receive service from the cells separated by at most a distance of $R_{\rm th}$, the cells that contribute into the load factor are those lying in the overlapping area $A_{\rm o}(r)$ and the user exclusion area $A_{\rm e}(r)$, as shown in Fig.~\ref{fig:subnetwork}\subref{fig:oneUE}. These areas can be expressed parametrically as follows
\begin{align} 
& A_{\rm o}(r) = 2r^2 - \theta + \frac{1}{2} \sin (2\theta), \label{eq:overlapping_area_2circle} \\
& A_{\rm e}(r) = \pi R_{\rm th}^2 - A_{\rm o}(r), \label{eq:nonoverlapping_user_area_2circle}
\end{align}
where $\theta \,{=}\, \cos^{{-}1}\left(\frac{r}{2R_{\rm th}}\right)$ is also depicted in Fig.~\ref{fig:subnetwork}\subref{fig:oneUE}.

The conditional load factor involved in (\ref{eqn:1st_moment_6}) could be expressed as follows
\begin{align}
\label{eqn:1st_moment_7}
\mathbb{E}[w |n_{\rm c} = i]& = \int_{0}^{R_{\rm th}}\mathbb{E}[w|r,n_{\rm c} = i]f_{r}(r){\rm d}r,
\end{align}
where $f_{r}(r) = 2r / R_{\rm th}$. The average load factor in (\ref{eqn:1st_moment_7}), which is conditioned on the distance $r$ and the number of cells $i$ (i.e., located within a circle of radius $R_{\rm th}$ around the origin), can be expressed as a sum in the form of a binomial expansion as follows
\begin{align}\label{eqn:1st_moment_8}
\mathbb{E}[w|r,n_{\rm c} = i] &= \sum_{k=0}^{i} \binom{i}{k}  \mathbb{E} [w|r,n_{A_{\rm o}}(r) = k] {p_{A_{\rm o}}(r)}^k (1-p_{A_{\rm o}}(r))^{i-k},
\end{align}
where $n_{A_{\rm o}}(r)$ stands for the number of cells in the overlapping area $A_{\rm o}(r)$, and $p_{A_{\rm o}}(r)$ is the probability of an SBS being in $A_{\rm o}(r)$. Since SBSs are distributed uniformly, we have $p_{A_{\rm o}}(r) \,{=}\, A_{\rm o}(r) / \pi R_{\rm th}^2$. In addition, each term in the summation of (\ref{eqn:1st_moment_8}) considers a case in which $k$ SBSs exist in the overlapping area $A_{\rm o}(r)$ out of a total of $i$ SBSs off the origin by at most the distance $R_{\rm th}$. 

While computing the average load expression at the right side of (\ref{eqn:1st_moment_8}) by employing the definition given in (\ref{eqn:load_factor_def}), one should take into account $k$ SBSs from the overlapping area $A_{\rm o}(r)$, $v$ SBSs from the user exclusion area $A_{\rm e}(r)$, and the single cell located at the origin as follows
\begin{align}\label{eqn:1st_moment_9}
\mathbb{E}[w|r,n_{A_{\rm o}(r)} = k] & =  \sum_{v = 0}^{\infty} \frac{1}{k+v+1} \, P\left\lbrace n_{A_{\rm e}}(r) = v \right\rbrace \nonumber \\
& = \sum_{v = 0}^{\infty} \frac{ \left[\nu_e(r)\right]^v e^{-\nu_e(r)} }{(k+v+1)\, v!},
\end{align} 
where $n_{A_{\rm e}}(r)$ is the random variable representing the number of SBSs in the user exclusion area $A_{\rm e}(r)$, which follows the Poisson distribution with rate $\nu_{\rm e}(r) \,{=}\, \rho_{\rm c} A_{\rm e}(r) \,{=}\, \nu_{\rm c} \,{-}\, \rho_{\rm c} A_{\rm o}(r)$. Finally, employing (\ref{eqn:1st_moment_7})-(\ref{eqn:1st_moment_9}) and $f_{r}(r) \,{=}\, 2r / R_{\rm th}$ in (\ref{eqn:1st_moment_6}), the first moment of $L$ is obtained as follows:
\begin{align} \label{eqn:load_1stmoment} 
\mathbb{E}\left[ L \right] & = \frac{2\nu_{\rm u}}{R_{\rm th}}
\sum_{v = 0}^{\infty} \sum_{i=0}^{\infty} \sum_{k=0}^{i} 
\frac{\nu_{\rm c}^i e^{-\nu_{\rm c}}}{(k{+}v{+}1)i!v!} \binom{i}{k} \int_{0}^{R_{\rm th}} 
\left[\nu_e(r)\right]^v e^{-\nu_e(r)} {p_{A_{\rm o}}(r)}^k (1{-}p_{A_{\rm o}}(r))^{i{-}k}
r{\rm d}r,
\end{align}
which is a function of the UE density $\nu_{\rm u}$, the SBS density $\nu_{\rm c}$, and the threshold distance $R_{\rm th}$.

\subsection{Second Moment of Load Variable}
Following the same approach of (\ref{eqn:1st_moment_1}), the second moment of $L$ can be written as
\begin{align}
\mathbb{E}\big[ L^2 \big]  &= \sum_{i=0}^{\infty}\sum_{j=0}^{\infty} \mathbb{E} \big[ L^2 | n_{\rm c} = i, n_{\rm u} = j \big]\, p_{\rm c}(i)\, p_{\rm u}(j), \nonumber\\
  &  =  \sum_{i=0}^{\infty}\sum_{j=0}^{\infty} \mathbb{E} \left[\left( \sum_{k=1}^{j} w_k \right)^2 \bigg|\, n_{\rm c} = i \right]p_{\rm c}(i)\, p_{\rm u}(j),
\label{eqn:2nd_moment_1}
\end{align}
which can be written after some manipulation as follows
\begin{align}\label{eqn:2nd_moment_2}
\mathbb{E} \left[ L^2 \right] & = \sum_{i=0}^{\infty} \sum_{j=0}^{\infty} \bigg( \sum_{k=1}^{j} \mathbb{E} \left[ w_k^2 | n_{\rm c} = i \right] + \sum_{k=1}^{j} \sum_{\substack{l=1 \\ l \neq k} }^{j} \mathbb{E} [w_k w_l |n_{\rm c} = i] \bigg) \, p_{\rm c}(i)\, p_{\rm u}(j)		\nonumber \\
& =  \sum_{i=0}^{\infty} \sum_{j=0}^{\infty} \Big( j \, \mathbb{E} \left[ w^2 | n_{\rm c} = i \right] + j (j-1) \, \mathbb{E} \left[w_k w_l | n_{\rm c} =i \right] \Big)  p_{\rm c}(i) p_{\rm u}(j)
\end{align}
for any $k,l$ with $k \neq l$. Following the discussion in obtaining (\ref{eqn:1st_moment_6}) from (\ref{eqn:1st_moment_5}), and employing first and second-order statistics of the Poisson distribution, we have
\begin{align}
\sum_{j=1}^{\infty} \, j (j-1) \, p_{\rm u}(j) & = \mathbb{E} \left[ n_{\rm u}^2 \right] - \mathbb{E} \left[ n_{\rm u} \right] = {\nu_{\rm u}}^2,
\label{eqn:2nd_moment_3}
\end{align}
and (\ref{eqn:2nd_moment_2}) accordingly becomes
\begin{align}\label{eqn:2nd_moment_4}
\mathbb{E} \left[ L^2 \right] &= \underbrace{\nu_{\rm u} \sum_{i=0}^{\infty} \mathbb{E} \left[ w^2 | n_{\rm c} = i \right] p_{\rm c}(i)}_{E_1} + 
\underbrace{ \nu_{\rm u}^2 \sum_{i=0}^{\infty} \mathbb{E} \left[w_k w_l | n_{\rm c} =i \right] p_{\rm c}(i)}_{E_2} .
\end{align}

The expectation $E[w^2 |n_{\rm c} = i]$ in \eqref{eqn:2nd_moment_4} can be computed following the steps of (\ref{eqn:1st_moment_7})-(\ref{eqn:1st_moment_9}) together with the modified version of (\ref{eqn:1st_moment_9}) given as
\begin{align}\label{eqn:2nd_moment_5}
\mathbb{E}[w^2|r,n_{A_{\rm o}}(r) = k] & =  \sum_{v = 0}^{\infty} \frac{ \left[\nu_{\rm e}(r)\right]^v e^{-\nu_{\rm e}(r)} }{(k+v+1)^2 \, v!} ,
\end{align}
and the first expression at the right hand side of \eqref{eqn:2nd_moment_4} becomes
\begin{align}\label{eqn:2nd_moment_e1}
\hspace{-0.0in} E_1 = \frac{2\nu_{\rm u}}{R_{\rm th}}
\sum_{v = 0}^{\infty} \sum_{i=0}^{\infty} \sum_{k=0}^{i} 
\frac{\nu_{\rm c}^i e^{-\nu_{\rm c}}}{(k{+}v{+}1)^2 i!v!} \binom{i}{k} \int_{0}^{R_{\rm th}} 
\left[\nu_e(r)\right]^v e^{-\nu_e(r)} {p_{A_{\rm o}}(r)}^k (1-p_{A_{\rm o}}(r))^{i-k}
r{\rm d}r.
\end{align}
However, computation of the second expectation in (\ref{eqn:2nd_moment_4}) is cumbersome due to the correlation between the individual load factors $w_k$ and $w_l$.

Because the expectation $\mathbb{E}[w_k w_l | n_{\rm c} = i]$ requires a second-degree analysis, we modify Fig.~\ref{fig:subnetwork}\subref{fig:oneUE} by adding a second user, and obtain Fig.~\ref{fig:subnetwork}\subref{fig:twoUE}. This new coordinate system has a SBS located at the origin, as before, and two UEs off this cell by random distances $r_{1}$ and $r_{2}$, both of which have the common distribution with $f_r(r) \,{=}\, 2r/R_{\rm th}$. We may have various orientations for relative positions of two UEs in Fig.~\ref{fig:subnetwork}\subref{fig:twoUE}, and therefore introduce a new variable $\omega$ which describes the difference of user angles with respect to the origin. 

Note that $\omega$ is actually the difference of two uniform random variables distributed between 0 and $2\pi$. The distribution of $\omega$ is therefore given as \cite{milios2009probability}
\begin{align}
\label{eqn:2nd_moment_6}
g(\omega) = 
\begin{cases}
	\displaystyle\frac{\omega}{2\pi ( 2\pi + 1 )}  & \mbox{if } \text{$ \omega \in [{-}2\pi,0]$}, \\
	\displaystyle\frac{1 -\omega}{4\pi^2}   & \mbox{if } \text{$ \omega \in [0,{+}2\pi]$}.
\end{cases}
\end{align}
The second-order expectation of interest could be accordingly written as
\begin{align}
\mathbb{E} [w_k w_l | n_{\rm c} = i]  & =  \int\displaylimits_{0}^{R_{\rm th}} \int\displaylimits_{0}^{R_{\rm th}} \int\displaylimits_{{-}2\pi}^{2\pi} E\left[w_k w_l | \, \textit{\textbf{r}}, \omega, n_{\rm c} = i \right] f_r(r_1) \, f_r(r_2) \, g(\omega) \, \mbox{d}\omega \, \mbox{d}r_1 \, \mbox{d}r_2, \label{eqn:2nd_moment_7}
\end{align}
which is counterpart of (\ref{eqn:1st_moment_7}) in the first moment computation, and where $\textit{\textbf{r}} = [r_1 \,\, r_2]$. 

\begin{figure}[!t]
\centering
\hspace*{-0.0in}
\subfloat[Case-I] {\includegraphics[width=0.28\textwidth]{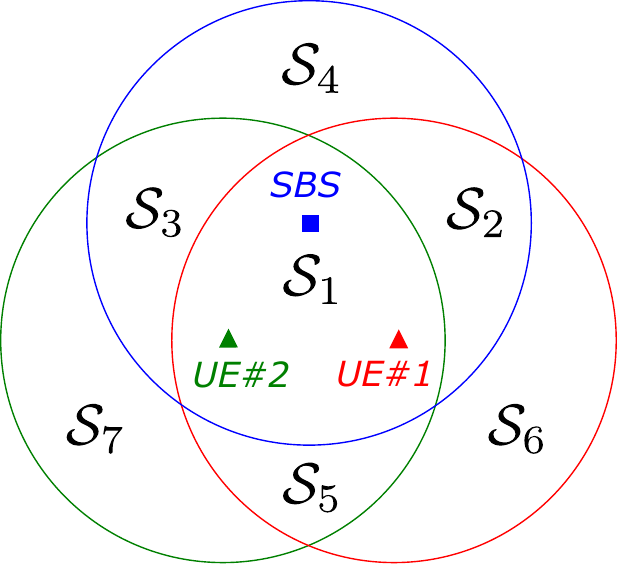} 		\label{fig:case1}}\vspace{0.0in}
\subfloat[Case-II] {\includegraphics[width=0.32\textwidth]{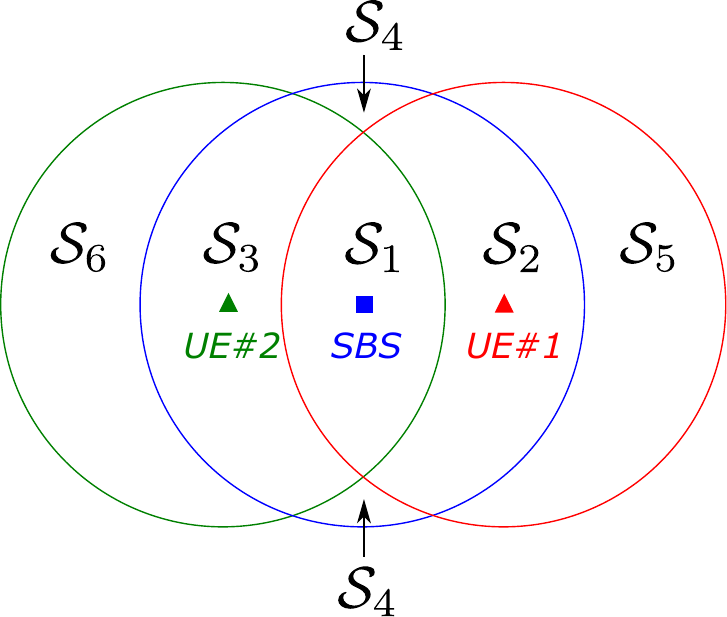} 		\label{fig:case2}}\vspace{0.0in}
\subfloat[Case-III] {\includegraphics[width=0.32\textwidth]{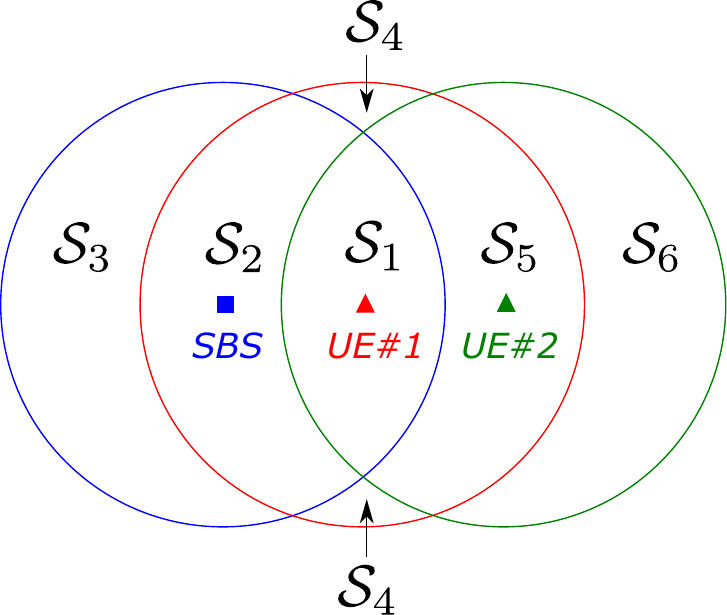} 		\label{fig:case3}}\vspace{-0.0in}
\caption{Relative orientations of two UEs around a single SBS.}
\label{fig:cases}
\end{figure}

In order to compute the expectation at the right side of (\ref{eqn:2nd_moment_7}), we need to consider various geometric orientations of two UEs around the origin, as in Fig.~\ref{fig:cases}. Among them, Case-I has a circular triangular overlapping area whereas Case-II and Case-III specify non-triangular overlapping areas. While the condition for the existence of a circular triangle area and respective area formulations are given in \cite{fewell2006area}, the non-triangular areas should be computed by employing (\ref{eq:overlapping_area_2circle}).

In order to express the term $E[ w_k w_l | \, \textit{\textbf{r}},\omega, n_{\rm c} = i] $ in the form of multinomial expansion, we need to take into account the number of constituent areas (i.e., $N$) forming the circular area of radius $R_{\rm th}$ around the origin (i.e., where the SBS is located). Note that the expectation in \eqref{eqn:2nd_moment_7} assumes $i\,{+}\,1$ SBSs in this circular region. Indeed, $N$ is a function of the angle $\omega$ given in Fig.~\ref{fig:subnetwork}\subref{fig:twoUE}, and all 3 cases sketched in Fig.~\ref{fig:cases} occurs for a certain set of $\omega$ values~\cite{fewell2006area}. Based on these 3 orientations in Fig.~\ref{fig:cases}, Case-I and Case-II have $N \,{=}\, 4$ constituent areas while Case-III has $N \,{=}\, 3$. As a counterpart of (\ref{eqn:1st_moment_8}), the desired expansion could therefore be given as
\begin{align} \label{eqn:2nd_moment_8}
E[ w_k w_l | \, \textit{\textbf{r}},\omega, n_{\rm c} = i] = \sum_{m_1=0}^{i} \hspace{0.1in} \dots \sum_{m_{N-1}=0}^{\substack{i-\sum\limits_{v=0}^{N-2} m_v}} 
E[w_k w_l | \, \textit{\textbf{r}}, \omega, \textit{\textbf{n}}(\textit{\textbf{r}},w) = \textit{\textbf{m}}] f(\textit{\textbf{m}};\textit{\textbf{p}}(\textit{\textbf{r}},w)),
\end{align}
where $\textit{\textbf{n}}(\textit{\textbf{r}},w)$ is the vector of the number of SBSs in each of the constituent areas, $\textit{\textbf{p}}(\textit{\textbf{r}},w)$ is the vector of multinomial probabilities associated with each of these areas, and $\textit{\textbf{m}}$ is the vector of summation indices. Each term of the summation in (\ref{eqn:2nd_moment_8}) corresponds to a unique distribution of the total of $i$ SBSs over the constituent areas. Specifically, the number of SBSs in the constituent area $A_v(\textit{\textbf{r}},w)$ is $n_v(\textit{\textbf{r}},w) = m_v$ for $v = 1,2,\dots, N$ with $\sum_{v=1}^{N} m_v = i$.

The probability mass function (PMF) in (\ref{eqn:2nd_moment_8}) is given as
\begin{align}\label{eqn:2nd_moment_9}
f\big(\textit{\textbf{m}};\textit{\textbf{p}}(\textit{\textbf{r}},w)\big) & = i! \, \prod_{v=1}^{N} (m_v!)^{-1} \prod_{v=1}^{N} {p_{A_v}(\textit{\textbf{r}},w)}^{m_v},
\end{align}
where $p_{A_v}(\textit{\textbf{r}},w)$ is the individual probability entry of $\textit{\textbf{p}}(\textit{\textbf{r}},w)$ associated with the constituent area $A_v(\textit{\textbf{r}},w)$, and is therefore given to be $p_{A_v}(\textit{\textbf{r}},w) = A_v(\textit{\textbf{r}},w) / \pi R_{\rm th}^2$ owing to the uniform distribution of SBSs in space. {\color{black}Note that $m_v$ SBSs in $A_v(\textit{\textbf{r}},w)$ can be placed in $m_v!$ different ways, and this makes $\prod_{v=1}^{N} m_v!$ considering all constituent areas. Since the total of $i$ SBSs can be ordered in $i!$ different ways, $i! \, \prod_{v=1}^{N} (m_v!)^{-1}$ in \eqref{eqn:2nd_moment_9} takes into account all possible relative SBS placements.}  

Following the philosophy behind (\ref{eqn:1st_moment_9}), and employing the PMF in \eqref{eqn:2nd_moment_9}, the expectation in the summation of (\ref{eqn:2nd_moment_8}) can be computed as follows
\begin{align}\label{eqn:2nd_moment_10}
E[w_k w_l | \, \textit{\textbf{r}}, \omega, \textit{\textbf{n}}(\textit{\textbf{r}},w)] &= \sum_{v_1 = 0}^{\infty} \sum_{v_2 = 0}^{\infty} \sum_{v_{\rm c} = 0}^{\infty}
P\left\lbrace n_{\rm e,c}(\textit{\textbf{r}},w) = v_{\rm c} \right\rbrace \prod_{s = 1}^{2} \frac{ P\left\lbrace n_{{\rm e},s}(\textit{\textbf{r}},w) = v_s \right\rbrace}{ n_{{\rm o},s}(\textit{\textbf{r}},w)+ v_s + v_{\rm c} + 1 },\\ 
&= \textcolor{black}{\sum_{v_1 = 0}^{\infty} \sum_{v_2 = 0}^{\infty} \sum_{v_{\rm c} = 0}^{\infty}
\frac{ \left[\nu_{\rm e,c}(r) \right]^{v_{\rm c}} e^{{-}\nu_{\rm e,c}(r)} }{v_{\rm c}!}
\prod_{s = 1}^{2} 
\frac{ \left[\nu_{{\rm e},s}(r) \right]^{v_{\rm s}} e^{{-}\nu_{{\rm e},s}(r) }  }{ v_{\rm s}! \left( n_{{\rm o},s}(\textit{\textbf{r}},w)+ v_s + v_{\rm c} + 1 \right) } }, \label{eqn:2nd_moment_11}
\end{align}
where $n_{\rm e,c}(\textit{\textbf{r}},w)$ and $n_{{\rm e},s}(\textit{\textbf{r}},w)$ are the number of SBSs in the common exclusion area $A_{\rm e,c}(\textit{\textbf{r}},w)$ and distinct exclusion area $A_{{\rm e},v}(\textit{\textbf{r}},w)$ for the $s$th UE, respectively, \textcolor{black}{which follow the Poisson distribution with rates $\nu_{\rm e,c}(r) \,{=}\, \rho_{\rm c} A_{\rm e,c}(r)$ and $\nu_{{\rm e},s}(r) \,{=}\, \rho_{\rm c} A_{{\rm e},s}(r)$, respectively,} with $s \,{=}\, 1,2$. We show all the exclusion and overlapping areas in Table~\ref{table} for the orientations considered in Fig.~\ref{fig:cases}.

\begin{table}[ht]
\renewcommand{\arraystretch}{1.1}\vspace{-2mm}
\caption{Overlapping and Exclusion Areas}
\label{table}
\centering
\begin{tabular}{c|c|c|c|}
\cline{2-4}
& Case-I    & Case-II   & Case-III	\\
\hline\hline
\multicolumn{1}{|r|}{$A_{\rm e,c}(\textit{\textbf{r}},w)$}      & $\mathcal{S}_5$                           & $\O$  				& $\mathcal{S}_5$ 			\\
\multicolumn{1}{|r|}{$A_{{\rm o},1}(\textit{\textbf{r}},w)$} 	& $\mathcal{S}_1 \bigcup \mathcal{S}_2$     & $\mathcal{S}_1 \bigcup \mathcal{S}_2$ 	& $\mathcal{S}_1 \bigcup \mathcal{S}_2$ \\
\multicolumn{1}{|r|}{$A_{{\rm e},1}(\textit{\textbf{r}},w)$} 	& $\mathcal{S}_6$ 		                    & $\mathcal{S}_5$ 	    & $\mathcal{S}_4$ 			\\
\multicolumn{1}{|r|}{$A_{{\rm o},2}(\textit{\textbf{r}},w)$} 	& $\mathcal{S}_1 \bigcup \mathcal{S}_3$     & $\mathcal{S}_1 \bigcup \mathcal{S}_3$		& $\mathcal{S}_1$ \\
\multicolumn{1}{|r|}{$A_{{\rm e},2}(\textit{\textbf{r}},w)$} 	& $\mathcal{S}_7$		                    & $\mathcal{S}_6$ 		& $\mathcal{S}_6$ \\
\hline
\end{tabular}
\end{table}

Note that $n_{{\rm o},s}(\textit{\textbf{r}},w)$ in \eqref{eqn:2nd_moment_10} is a given (i.e., deterministic) value representing the number of SBSs in the overlapping area $A_{{\rm o},s}(\textit{\textbf{r}},w)$, with $s\,{=}\,1,2$. More specifically, $n_{{\rm o},s}(\textit{\textbf{r}},w)$ is the sum of the entries of $\textit{\textbf{n}}(\textit{\textbf{r}},w)$ associated with the constituent areas forming $A_{{\rm o},s}(\textit{\textbf{r}},w)$, which are explicitly given in Table~\ref{table} for $s\,{=}\,1,2$. As an example, we have $n_{{\rm o},1}(\textit{\textbf{r}},w) = n_{1}(\textit{\textbf{r}},w)+ n_{2}(\textit{\textbf{r}},w)$ and $n_{{\rm o},2}(\textit{\textbf{r}},w) = n_{1}(\textit{\textbf{r}},w)+ n_{3}(\textit{\textbf{r}},w)$ for Case-I, where $n_i(\textit{\textbf{r}},w)$ is the number of SBSs in the area $\mathcal{S}_i$ for $i \,{=}\, 1,2,3$.

As a particular case, since $A_{\rm e,c}(\textit{\textbf{r}},w)$ does not exist for Case-II, (\ref{eqn:2nd_moment_11}) simplifies to
\begin{align}\label{eqn:2nd_moment_12}
E[w_k w_l | \, \textit{\textbf{r}}, \omega, \textit{\textbf{n}}(\textit{\textbf{r}},w)] 
= \sum_{v_1 = 0}^{\infty} \sum_{v_2 = 0}^{\infty}
\prod_{s = 1}^{2} \frac{ \left[\nu_{{\rm e},s}(r) \right]^{v_{\rm s}} e^{{-}\nu_{{\rm e},s}(r)}  }{ v_{\rm s}! \left( n_{{\rm o},s}(\textit{\textbf{r}},w)+ v_s + 1 \right) }.
\end{align}
Combining \eqref{eqn:2nd_moment_7}-\eqref{eqn:2nd_moment_11}, we finally obtain $E_2$ appearing in \eqref{eqn:2nd_moment_4} as follows
\begin{align} \label{eqn:2nd_moment_e2}
&\hspace{0.0in}E_2 = \frac{4\nu_{\rm u}^2}{R_{\rm th}^2} \sum_{i=0}^{\infty} \sum_{v_1 = 0}^{\infty} \sum_{v_2 = 0}^{\infty} \sum_{v_{\rm c} = 0}^{\infty} \nu_{\rm c}^i e^{-\nu_{\rm c}} \!\!
\int\displaylimits_{0}^{R_{\rm th}} \int\displaylimits_{0}^{R_{\rm th}} \int\displaylimits_{{-}2\pi}^{2\pi} 
\sum_{m_1=0}^{i} \hspace{0.1in} \hspace{-0.1in} \dots \!\!\! \sum_{m_{N-1}=0}^{\substack{i-\sum\limits_{v=0}^{N-2} m_v}} \!\!\!
\frac{ \left[\nu_{\rm e,c}(r) \right]^{v_{\rm c}} e^{{-}\nu_{\rm e,c}(r)} }{v_{\rm c}! \, m_1!\dots m_N!} \prod_{v=1}^{N} {p_{A_v}(\textit{\textbf{r}},w)}^{m_v} \nonumber \\
&\hspace{2in} \times \prod_{s = 1}^{2} 
\frac{ \left[\nu_{{\rm e},s}(r) \right]^{v_{\rm s}} e^{{-}\nu_{{\rm e},s}(r)}  }{ v_{\rm s}! \left( n_{{\rm o},s}(\textit{\textbf{r}},w) {+} v_s {+} v_{\rm c} {+} 1 \right) } \, g(\omega) \, r_1 r_2 \, \mbox{d}\omega \,\mbox{d}r_1 \mbox{d}r_2,
\end{align}
which is also a function of densities $\nu_{\rm u}$ and $\nu_{\rm c}$, and the distance $R_{\rm th}$.

As a result, the respective parameters $\alpha$ and $\beta$ of the fitting gamma distribution can be computed using the first order moment $\mathbb{E}[L]$ given in \eqref{eqn:load_1stmoment}, and the second order moment $\mathbb{E}[L^2]$ given in \eqref{eqn:2nd_moment_4} (i.e., the sum of \eqref{eqn:2nd_moment_e1} and \eqref{eqn:2nd_moment_e2}), based on the relations given in \eqref{eqn:gamma_alpha} and \eqref{eqn:gamma_beta}. The CDF of load distribution can therefore be written as
\begin{align}
F_L(x) = P\{L< x \} = e^{-\nu_{\rm u}} +(1-e^{-\nu_{\rm u}})\int_{0^+}^{x}\frac{\beta^{\alpha}}{\Gamma(\alpha)} y^{\alpha -1}e^{-\beta y}{\rm d}y,
\label{eqn:load_cdf}
\end{align}
where first term represents the void probability, $P\{L \,{=}\, 0\}$ (i.e., no user is around the SBS). 
Using \eqref{eqn:load_cdf}, the respective PDF of load distribution can be written as
\begin{align}
\label{eqn:load_pdf}
f_{L}(x) = 
\begin{cases}
	\displaystyle e^{-\nu_{\rm u}}  & \mbox{if } \text{$x = 0$}, \\
	\displaystyle  (1 - e^{-\nu_{\rm u}}) \frac{\beta^{\alpha}}{\Gamma(\alpha)} x^{\alpha -1}e^{-\beta x} & \mbox{if } \text{$x > 0$}.
\end{cases}
\end{align}

\section{Load Based On/Off Scheduling} \label{sec:algorithms}

In this section, we study OOS strategies with a goal of having more energy-efficient SCNs. In this respect, we first consider a random OOS algorithm (i.e., ROO) to set up a simple benchmark to evaluate performance of smarter OOS strategies. We then propose two novel load based OOS algorithms, which are called CLB and DLB, and establish a good compromise between energy-efficiency and network throughput. Finally, we also consider a more sophisticated OOS strategy, which is called WUC, where the macrocell has the full capability to wake up any sleeping SBSs.

We assume that the percentage of sleeping SBSs
are fixed in all the OOS algorithms under consideration for the sake of a fair comparison. As a result, for each sleeping SBS to wake up, the OOS algorithms choose the \textit{best idle} SBS to turn off. Depending on the specific OOS algorithm, the set of sleeping SBSs may dynamically change as the turn-off and turn-on events occur repeatedly. 

We also assume that any UE can get service from the available SBSs, which are either \textit{currently idle} or \textit{become idle} within the waiting time period, as discussed in Section~\ref{sec:scn_model}. In particular, ROO, CLB, and DLB strategies assume no capability at the central controller to wake up a sleeping SBS during its random sleep time. The WUC strategy, however, assumes that the central controller can give order to wake up a sleeping SBS to make it available within the waiting time (i.e., which would otherwise not become available).


\subsection{Random On/Off Scheduling} \label{sec:algorithm_roo}
In this strategy, a central controller (e.g., macrocell) determines which SBS to turn off randomly, and assigns a random sleep time for each SBS having turned off. Each sleeping SBS wakes up automatically after its sleep time expires, and the central controller decides which SBS to turn off in return.
The overall procedure is given in Algorithm~\ref{alg:roo}.
\begin{algorithm}
\caption{Random On/Off Scheduling}
\begin{algorithmic}[1]
\State \textbf{Input}: The sleep time of $i$th SBS has expired
\State $\text{SBS}_{\rm nextToSleep} \gets \text{ROO}(i,\mathcal{S}_{\rm all})$ \Comment{$\mathcal{S}_{\rm all}$ is the set of all SBSs}
\State turn off $\text{SBS}_{\rm nextToSleep}$ 
\Procedure{ROO}{$i$,$\mathcal{S}_{\rm all}$} \Comment{ROO algorithm}
\State $\mathcal{S}_{\rm idle} \gets$ $\find_{1 \leq \ell \leq \left|\mathcal{S}_{\rm all}\right|}(\,\mathrm{state}(\mathcal{S}_{\rm all}(\ell))==\text{idle}\, )$ 
	\State $j \gets \mathrm{rand} (1,\left|\mathcal{S}_{\rm idle}\right|)$
\State \Return $\mathcal{S}_{\rm idle}(j)$
\EndProcedure
\end{algorithmic}\label{alg:roo}
\end{algorithm}


\subsection{Centralized Load Based On/Off Scheduling}\label{sec:algorithm_clb}
The CLB can be considered to be the load based alternative of ROO, which operates in a centralized fashion as described in Algorithm~\ref{alg:clb}. In CLB, the central controller turns off the SBS with the minimum instantaneous load value computed using~\eqref{eqn:load_def} as a response to each SBS that has just waken up.
Note that the algorithm needs the load values of idle and only; however, UE shares its instantaneous load factor with the idle and active cells because of two reasons: $i)$ each active cell may return to idle status after completion of the transmission, therefore, may be available within $w_t$ time, $ii)$ density of non-sleeping cells (i.e. idle and active) do not change therefore, distribution of load is can be obtained, which allows implemetation of on/off decision in distributed manner. 
\begin{algorithm}
\caption{Centralized Load Based On/Off Scheduling}
\begin{algorithmic}[1]
\State \textbf{Input}: The sleep time of $i$th SBS has expired
\State $\text{SBS}_{\rm nextToSleep} \gets \text{CLB}(i,\mathcal{S}_{\rm all})$ \Comment{$\mathcal{S}_{\rm all}$ is the set of all SBSs}
\State turn off $\text{SBS}_{\rm nextToSleep}$ 
\Procedure{CLB}{$i$,$\mathcal{S}_{\rm all}$} \Comment{CLB algorithm}
\State $\mathcal{S}_{\rm idle} \gets$ $\find_{1 \leq \ell \leq \left|\mathcal{S}_{\rm all}\right|}(\,\mathrm{state}(\mathcal{S}_{\rm all}(\ell))==\text{idle}\, )$ 
    \State compute $L_\ell$ by \eqref{eqn:load_def} for $\ell = 1, \dots, \left|\mathcal{S}_{\rm idle}\right|$ 
	\State $j \gets \argmin_{1 \leq \ell \leq \left|\mathcal{S}_{\rm idle}\right|} L_\ell$ 
\State \Return $\mathcal{S}_{\rm idle}(j)$
\EndProcedure
\end{algorithmic}\label{alg:clb}
\end{algorithm}


\subsection{Distributed Load Based On/Off Scheduling} \label{sec:algorithm_dlb}

The DLB algorithm is a distributed version of the centralized CLB algorithm, where the overall operation does not need a central controller. In DLB approach, whenever a sleeping SBS is to wake up (i.e., after expiration of its random sleep time), that specific SBS is designated to be the decision-maker to decide the next SBS to be turned off. The decision-maker SBS first determines its all idle first-hop neighbours (i.e., within a distance of at most $R_{\rm th}$) as the candidate SBSs to be turned off. The instantaneous load values of the candidate SBSs are then collected (e.g., via BS-BS communication using X2 backhaul link~\cite{blogowski2012dimensioning}), and the one with the minimum instantaneous load is chosen by the decision-maker SBS as the one to turn off next. 

An important feature of DLB is the mechanism specifying when to stop searching candidates in a \textit{wider} neighborhood. To this end, the algorithm checks the following relation
\begin{align}\label{eqn:dlb_check}
    1-(1- P\{L<L_{\rm min} \})^{|\mathrm{S}(k+1)|} < \kappa
\end{align}
where $L_{\min}$ is the minimum instantaneous load associated among the cells traversed up to $k$ hops, and $\kappa$ is a threshold probability. Given the cardinality of $|S(k+1)|$ idle cells at next hop, $k+1$, \eqref{eqn:dlb_check} checks the probability of finding a cell with a lower load than that of $L_{\min}$.  Note that \eqref{eqn:dlb_check} can be computed readily using the analytical load CDF in~\eqref{eqn:load_cdf}. If inequality of \eqref{eqn:dlb_check} is correct, then the algorithm stops searching for a better candidate SBS, and decides to turn off the current candidate. Otherwise, the algorithm widens its search to second-hop neighbours (i.e., those in $2R_{\rm th}$  distance). Likewise, algorithm continues to widen its search till it becomes less likely to find an SBS with a lower load than that of the existing candidate (i.e., for which \eqref{eqn:dlb_check} turns out to be true). The complete procedure is given in Algorithm~\ref{alg:dlb}.


\begin{algorithm}
\caption{Distributed Load Based On/Off Scheduling}
\begin{algorithmic}[1]
\State \textbf{Input}: The sleep time of $i$th SBS has expired
\State $\text{SBS}_{\rm nextToSleep} \gets \text{DLB}(i,\kappa)$
\State turn off $\text{SBS}_{\rm nextToSleep}$ 
\Procedure{DLB}{$i$,$\kappa$} \Comment{DLB algorithm}
\State $L_{\min} \gets \infty$, $k \gets 1$
\While{ $1-(1- F_L\left(L_{\min}\right) \})^{|\mathrm{S}(k+1)|} > \kappa$}

\State $\mathcal{S} \gets$ $\find_{1 \leq \ell \leq \left|\mathcal{S}_{\rm all}\right|}( \, \mathrm{distance}(\mathcal{S}_{\rm all}(\ell),\mathcal{S}_{\rm all}(i)) \leq k R_{\rm th} \,)$ \Comment{$\mathcal{S}_{\rm all}$ is the set of all SBSs}
\State $\mathcal{S}_{\rm idle} \gets$ $\find_{1 \leq \ell \leq \left|\mathcal{S}\right|}(\,\mathrm{state}(\mathcal{S}(\ell))==\text{idle}\, )$ 
    \State compute $L_\ell$ by \eqref{eqn:load_def} for $\ell = 1, \dots, \left|\mathcal{S}_{\rm idle}\right|$ 
	\State $j \gets \argmin_{1 \leq \ell \leq \left|\mathcal{S}_{\rm idle}\right|} L_\ell$ 
	\State $L_{\min} = L_j$
\State $k \gets k + 1$
\EndWhile
\State \Return $\mathcal{S}_{\rm idle}(j)$
\EndProcedure
\end{algorithmic}\label{alg:dlb}
\end{algorithm}

\subsection{Wake-up Control Based On/Off Scheduling} \label{sec:algorithm_wuc}

We finally consider a more complex approach, which is called wake-up control (WUC) and given in Algorithm~\ref{alg:wuc}. This algorithm is, indeed, very similar to the CLB algorithm, except that the central controller now has the full control to wake up any sleeping SBS (even before the respective sleep time expires). By this way, any of the UE service requests, which could not otherwise be met by available idle SBSs, might be handled by incorporating the sleeping SBSs. To do so, the candidate sleeping SBSs should be within the communication range, and be able to wake up within the tolerable delay of that UE holding the current request. More specifically, the boot-up time of the candidate sleeping SBSs (i.e., given in Table~\ref{tab:sbs_states}) should end within the tolerable delay. Note that once the central controller places a wake-up order for the nearest candidate SBS, it is classified as \textit{reserved} to avoid from placing another wake-up order for the same SBS (for another UE request). Although this approach decreases the blocking probability of SCN, the energy consumption is likely to increase since sleeping SBSs getting wake-up orders cannot remain in their low-power consumption states.      

\begin{algorithm}
\caption{WUC Based Service Request Handling}
\begin{algorithmic}[1]
\State \textbf{Input}: UE service request arrival at $t_{\rm now}$ \Comment{$t_{\rm now}$ is the current time}
\State $t_{\rm deadline} \gets t_{\rm now} + w_{\rm t} $
\While{ $t_{\rm now} \leq t_{\rm deadline}$}
\State $\text{SBS}_{\rm best} = \text{WUC}(t_{\rm deadline},\,t_{\rm now},\mathcal{S}_{\rm all})$ \Comment{$\mathcal{S}_{\rm all}$ is the set of all SBSs}
\State update $t_{\rm now}$
\EndWhile
\If{ $\text{SBS}_{\rm best} == \varnothing $}
\State service request is blocked
\Else
\State associate UE to $\text{SBS}_{\rm best}$
\EndIf
\Procedure{WUC}{$t_{\rm deadline}$,\,$t$} \Comment{WUC algorithm}
\State $\mathcal{S} \gets$ $\find_{1 \leq \ell \leq \left|\mathcal{S}_{\rm all}\right|}\,(\, \mathrm{distance}(\mathcal{S}_{\rm all}(\ell),\text{UE} \,) \leq R_{\rm th} \,)$
\State $\mathcal{S}_{\rm candidate} \gets$ $\find_{1 \leq \ell \leq \left|\mathcal{S}\right|}\,(\,t + \mathrm{bootupTime}(\mathcal{S(\ell)}) \leq t_{\rm deadline}\, )$ 
\If{$\mathcal{S}_{\rm candidate} == \varnothing$}
\State \Return $\varnothing$
\Else
\State $j \gets \argmin_{1 \leq \ell \leq \left|\mathcal{S}_{\rm candidate}\right|} \mathrm{distance}(\mathcal{S}_{\rm candidate}(\ell),\text{UE} \,)$ 
\State \Return $\mathcal{S}_{\rm candidate}(j)$
\EndIf
\EndProcedure
\end{algorithmic}\label{alg:wuc}
\end{algorithm}

\section{Simulation Results} \label{sec:simulations}

In this section, we present numerical results for the performance of i) proposed load definition in representing the actual traffic load of SCN, and ii) novel load based OOS strategies. In particular, performance of the novel CLB and DLB algorithms are evaluated in comparison to the ROO and WUC algorithms as the benchmark OOS strategies, and the static topology without dynamic OOS approach. 

We assume a circular area with a radius of $250\,\text{m}$ for the deployment of UEs and SBSs, and the results are averaged over $1000$ iterations and $10000$ seconds of simulation time. In terms of overall SCN traffic, we consider two main scenarios: low network utilization ($1\%$) and (relatively) high network utilization ($20\%$). In both scenarios, UE traffic profile (i.e., service request rate and associated file size) is assumed to be adequate so that there is enough room to effectively apply OOS strategies (i.e, all SBSs would otherwise be occupied all the time). For delayed access scheme, we assume a sufficiently large but reasonable UE delay tolerance of $60$ sec (as well as zero tolerable delay), which enables WUC algorithm to attain its best performance, and, hence, the performance gap between WUC and other strategies becomes apparent. All the simulation parameters are listed in Table~\ref{tab:simulation_parameters}.

\begin{table}[ht]
	\caption{Simulation Parameters}
	\label{tab:simulation_parameters}
	\centering
	\begin{tabular}{lc}
		\hline
		Parameter   & Value	\\
		\hline\hline
		Cell density $(\rho_{\rm c})$           & $0.0005$ m$^{{-}2}$ \\
		User density $(\rho_{\rm u})$ 		    & $0.0005$ m$^{{-}2}$ \\
		Service request rate $(\lambda_{\rm U})$   & $\{0.001,0.01\}$ s$^{-1}$	\\
		Average file size $(1/\lambda_{\rm F})$	& $\{1,2\}$ MB \\
		Sleep rate $(\lambda_{\rm S})$			& $\{0.001,0.002\}$ s$^{-1}$ \\
	 	Tolerable delay $(w_{\rm t})$			& $\{0,60\}$ s \\
	 	Threshold distance $(R_{\rm th})$ 		& $50$ m \\
		Bandwidth $(\mathsf{BW})$				& $1$ MHz \\
		Signal-to-Noise Ratio ($\mathsf{SNR}$) 	& $20$ dB \\
		Threshold probability for DLB $(\kappa)$ & $0.3$ \\
		Maximum search range for DLB  			& $3 \times R_{\rm th}$ \\
		Path loss exponent ($\alpha$)    		& $4$ \\
 		\hline
	\end{tabular}
\end{table}

\subsection{Performance Metrics}
In the performance analysis, we consider the following criteria. 

\begin{itemize}
	
	\item[-] \textit{Blocking Probability}: The fraction of rejected service requests among all, which is basically due to sleeping or fully occupied (i.e., actively transmitting) SBSs, which is given as
	\begin{align}\label{eqn:blocking_probability}
	\mathsf{P_{block}} = \frac{\text{number of rejected service requests}}{\text{total number of service requests}}.
	\end{align}
	
	\item[-] \textit{Average Throughput}: The total number of bits transmitted averaged over the total simulation time, which is also normalized with respect to the number of users as follows
	\begin{align}\label{eqn:average_throughput}
	\mathsf{R_{SCN}} = \frac{\text{total number of transmitted bits}}{\text{number of users} \times \text{simulation time}}\,\text{(bps).}
	\end{align}
	The number of transmitted bits in \eqref{eqn:average_throughput} is given by the Shannon capacity formula as follows 
	\begin{align}
	    \mathsf{R} = \mathsf{BW} \log_2 \left( 1 + \mathsf{SINR} \right),
	\end{align}
	where $\mathsf{BW}$ is the transmission bandwidth, and $\mathsf{SINR}$ is the signal-to-interference-plus-noise ratio. Assuming the association between $i$th UE and $j$th SBS, the respective $\mathsf{SINR}$ at the UE side is defined as follows
	\begin{align}
	    \mathsf{SINR}_{ij} = 
	    \frac{\displaystyle d_{ij}^{{-}\alpha} }
	    {\displaystyle \sum\limits_{\ell \neq j} d_{i\ell}^{{-}\alpha} +1/\mathsf{SNR}},
	\end{align}	
	where $d_{ij}$ is the distance between $i$th UE and $j$th active SBS, $\alpha$ is the path loss (PL) exponent, and $\mathsf{SNR}$ is the signal-to-noise ratio.
	
	\item[-] \textit{Normalized Energy Efficiency}: The amount of energy consumed for each transmitted bit averaged over the total simulation time, which is also normalized by the number of users and the maximum power $\mathsf{P_{max}}$ associated with the active state. 
	\begin{align}\label{eqn:energy_efficiency}
	\mathsf{EE} = \frac{ \mathsf{R_{SCN}} } {\text{total energy consumption}}\,\times \, \mathsf{P_{max}}\, \text{(bps/joule).}
	\end{align}
	Note that the power consumption of an SBS at each state is given in Table~\ref{tab:sbs_states} as the fraction of the maximum power $\mathsf{P_{max}}$, and we therefore use these power fractions while computing~\eqref{eqn:energy_efficiency}.
\end{itemize}

\subsection{Load Distribution Verification}

\begin{figure}[!t]
\centering\hspace{-0.25in}
\subfloat[CDF] {\includegraphics[width=0.52\textwidth]{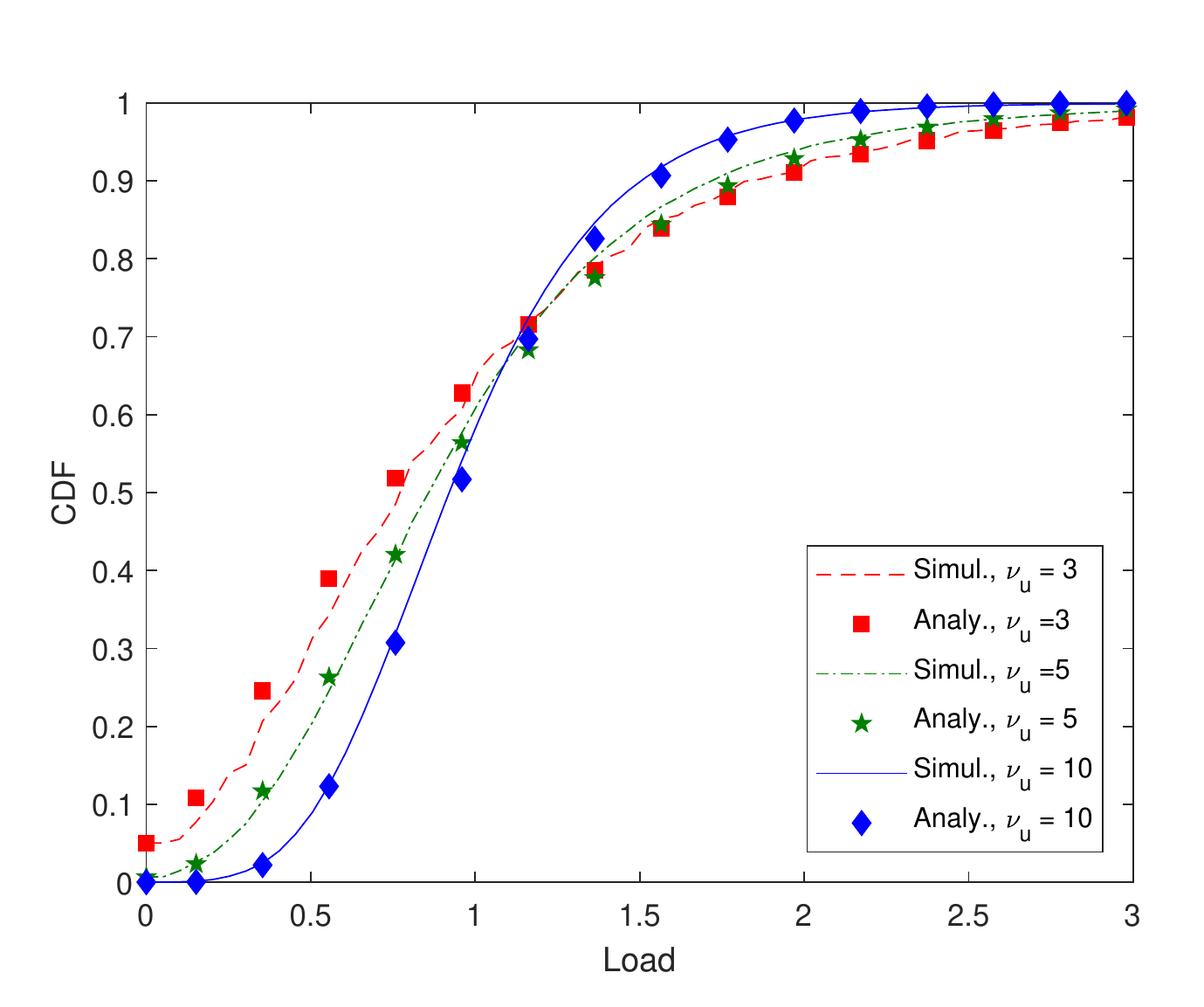}
\label{fig:cdf}}\hspace{-0.25in}
\subfloat[PDF] {\includegraphics[width=0.52\textwidth]{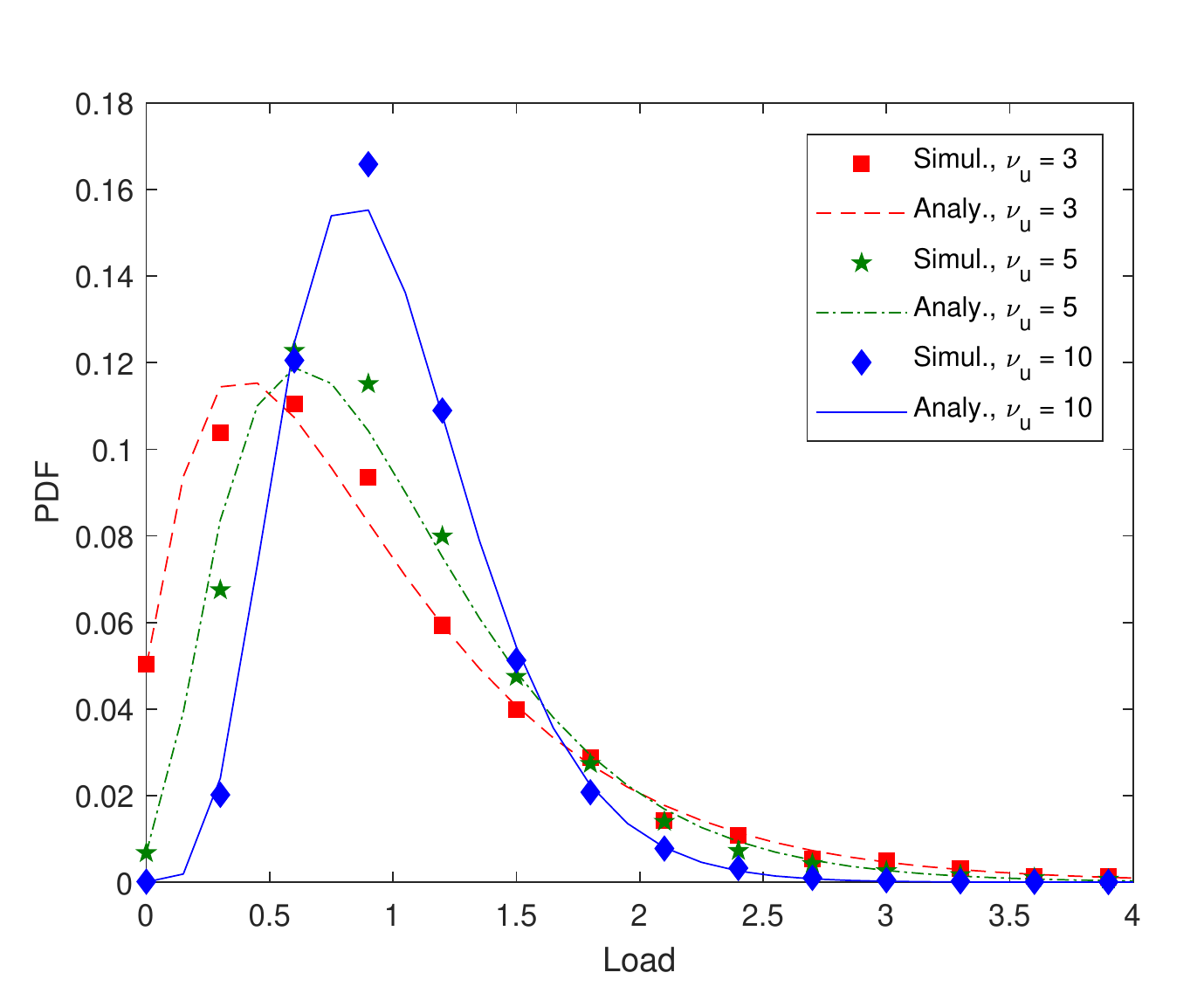}
\label{fig:pdf}}		
\caption{Analytical and simulation results for load distribution for range-dependent UE densities of $\nu_{\rm u} \,{=}\, \{3,5,10\}$ and $\rho_{\rm c} {/} \rho_{\rm u} \,{=}\, 1$.}
\label{fig:distribution}
\end{figure}

In Fig.~\ref{fig:distribution}, we depict the CDF and PDF of the SCN traffic load for range-dependent UE densities of $\nu_{\rm u} \,{=}\,\{3,5,10\}$, where extensive simulation results are provided along with the analytical results computed using \eqref{eqn:load_cdf}. We observe that the analytical results nicely match the simulations for all three UE densities, which verifies the respective derivation in Section~\ref{sec:derivations}. Accuracy of approximation depends on achievable precision of load values which ultimately depends the precision of load factor. Precision of load increment improves as load factor becomes small. Therefore, as we increase UE's range, load factor of UE becomes smaller, and the approximate load distribution converges to the exact load distribution. We, therefore, start observing finely matched load distribution in low load regime as the average number of UEs around cell increases. 



\subsection{Low Utilization Performance}

\begin{figure}[!t]
	\centering
    \includegraphics[width=0.7\textwidth]{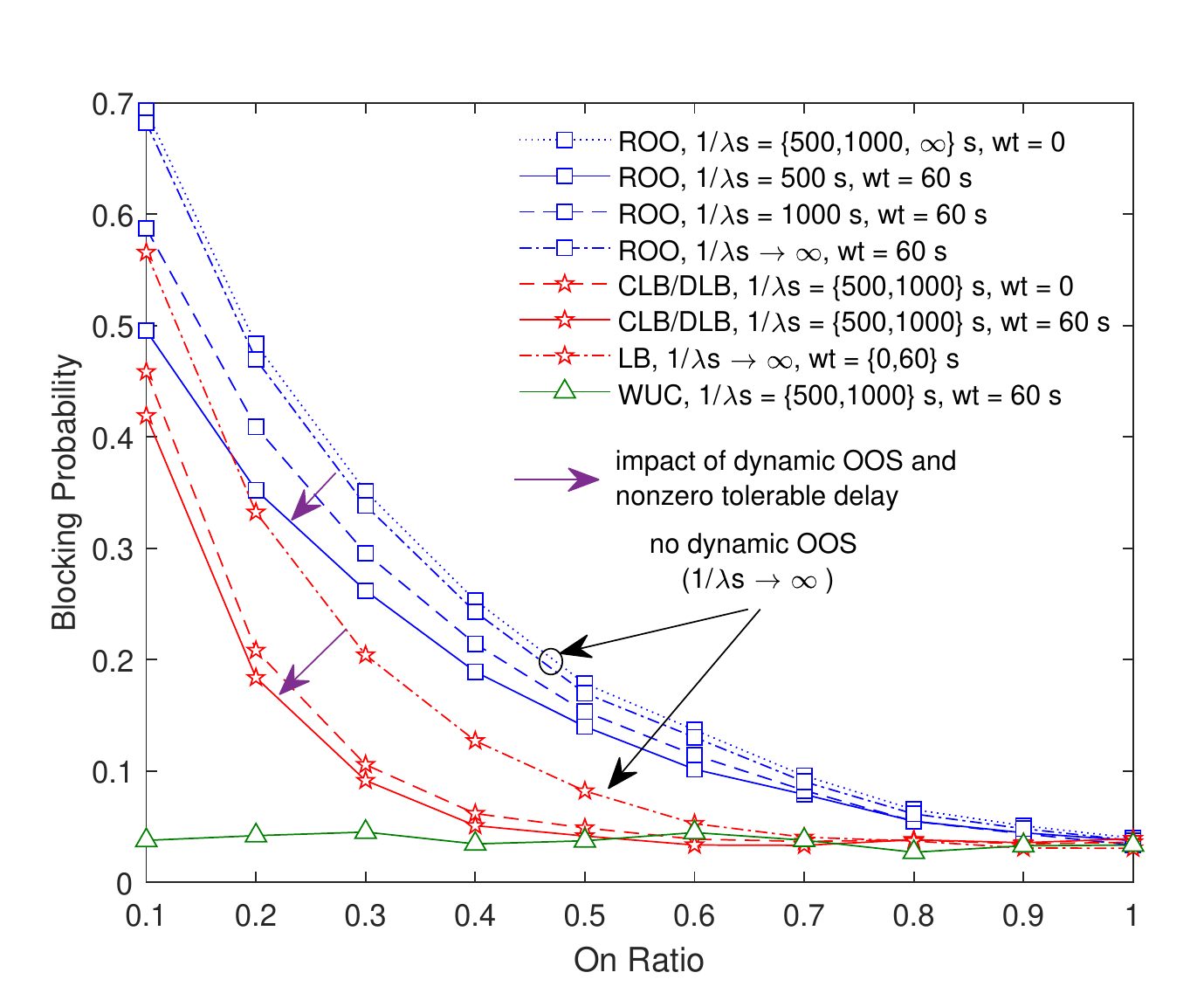}
	\caption{Blocking probability $\mathsf{P_{block}}$ along with on ratio (i.e., fraction of non-sleeping SBSs) for $1/\lambda_{\rm S}\,{\in}\,\{500\,\text{s}, 1000\,\text{s}, \infty\}$ and $w_{\rm t}\,{\in}\,\{0,60 \, \text{s}\}$ assuming low network utilization of $1\%$ (i.e., $1/\lambda_{\rm U} \,{=}\, 1000\,\text{s}$ and $1/\lambda_{\rm F} \,{=}\, 1\,\text{MB}$).}
	\label{fig:lowutil_pblock}
\end{figure}

In this subsection, we consider the performance of OOS strategies under a low network utilization scenario, where the UE service request rate and average file size are $1/\lambda_{\rm U} \,{=}\, 1000\,\text{s}$ and $1/\lambda_{\rm F} \,{=}\, 1\,\text{MB}$, respectively. Together with the UE and SBS densities given in Table~\ref{tab:simulation_parameters}, respective network utilization is on the order of $1\%$ based on the utilization results that we have left out due to the space limitations.
In Fig.~\ref{fig:lowutil_pblock}, we present blocking probability results for all the algorithms under consideration against varying \emph{on-ratio} (i.e. fraction of non-sleeping SBSs). In particular, we take into account the effect of sleep rate $\lambda_{\rm S}$ (or equivalently sleep period $1/\lambda_{\rm S}$) and waiting time $w_{\rm t}$ by assuming $1/\lambda_{\rm S}\,{\in}\,\{500\,\text{s}, 1000\,\text{s}, \infty\}$ and $w_{\rm t}\,{\in}\,\{0,60\,\text{s}\}$. Note that $1/\lambda_{\rm S}\,{\rightarrow}\,\infty$ corresponds to a scenario with no dynamic OOS events, i.e., topology of non-sleeping SBSs does not change once it is initialized at the beginning. We therefore describe the respective load based algorithm simply with LB since either centralized or distributed strategy (i.e., in CLB and DLB) is only applicable with dynamic on/off events occurring after initialization. 

We observe in Fig.~\ref{fig:lowutil_pblock} that blocking probabilities for any OOS algorithm decrease as either more SBSs become available (i.e., increasing on ratio), or tolerable delay gets larger (i.e., more room to meet UE service request). In particular, the load based CLB and DLB perform much better than the random scheme ROO in terms of achieving less blocking events (i.e., rejected UE requests). Note that CLB and DLB actually have the same performance for any choice of on ratio, and we therefore referred to this common performance as CLB/DLB. This equity underscores the power of DLB especially for large-scale SCNs in the sense that DLB does not need information of \textit{all} SBSs (i.e., in contrast to CLB) to decide the next SBS to turn off, and is hence more efficient to implement. Considering a wide range of reasonable non-sleeping SBS fractions (i.e., greater than $0.5$ for a realistic SCN), CLB/DLB is shown to attain the performance of more complex WUC scheme, where ROO still falls short of that level.

In Fig.~\ref{fig:lowutil_pblock}, the response of random and load based algorithms to the choice of sleep period $1/\lambda_{\rm S}$ and waiting time $w_{\rm t}$ are observed to have some interesting differences. Assuming zero tolerable delay (i.e., $w_{\rm t} \,{=}\, 0$), the blocking probability of random scheme ROO does not change at all along with $1/\lambda_{\rm S}$ even considering the no dynamic OOS case (i.e., $1/\lambda_{\rm S}\,{\rightarrow}\,\infty$). When we consider nonzero tolerable delay (i.e., $w_{\rm t} \,{=}\, 60\,\text{s}$), we start observing significant performance improvement in ROO along with decreasing $1/\lambda_{\rm S}$, where the best performance occurs at  $1/\lambda_{\rm S} \,{=}\, 500\,\text{s}$. On the other hand, load based CLB/DLB achieves significantly better performance for $1/\lambda_{\rm S} \,{=}\, \{500\,\text{s},1000\,\text{s}\}$ (as compared to no dynamic OOS case) even under zero tolerable delay condition. When a nonzero tolerable delay (i.e., $w_{\rm t} \,{=}\, 60$ s) is further assumed, the best performance is even superior to that of the zero tolerable delay, but the respective performance gap remains marginal. As a result, CLB/DLB is more robust to \textit{delay intolerance} while random scheme ROO requires \textit{longer tolerable delays} for performance improvement. In addition, applying OOS dynamically is useful for ROO only when the delay tolerance is sufficiently large, while dynamic OOS improves performance of CLB/DLB in both delay tolerant and intolerant SCNs.


\begin{figure}[!t]
\centering\hspace{-0.0in}
\subfloat[Average Throughput, $\mathsf{R_{SCN}}$] {\includegraphics[width=0.68\textwidth]{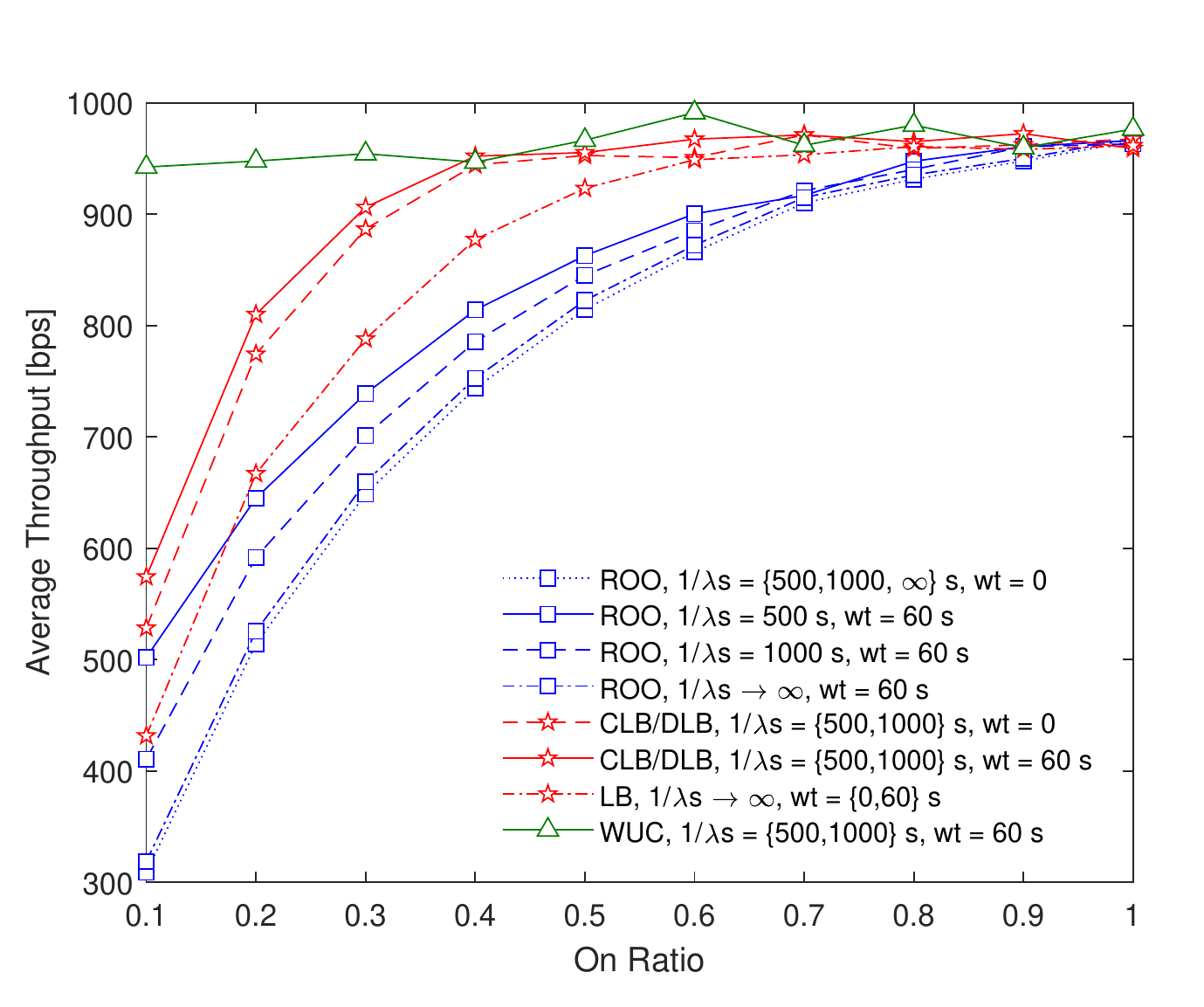}
\label{fig:lowutil_rate}}\vspace{-0.3in}
\subfloat[Normalized Energy Efficiency, $\mathsf{EE}$] {\includegraphics[width=0.68\textwidth]{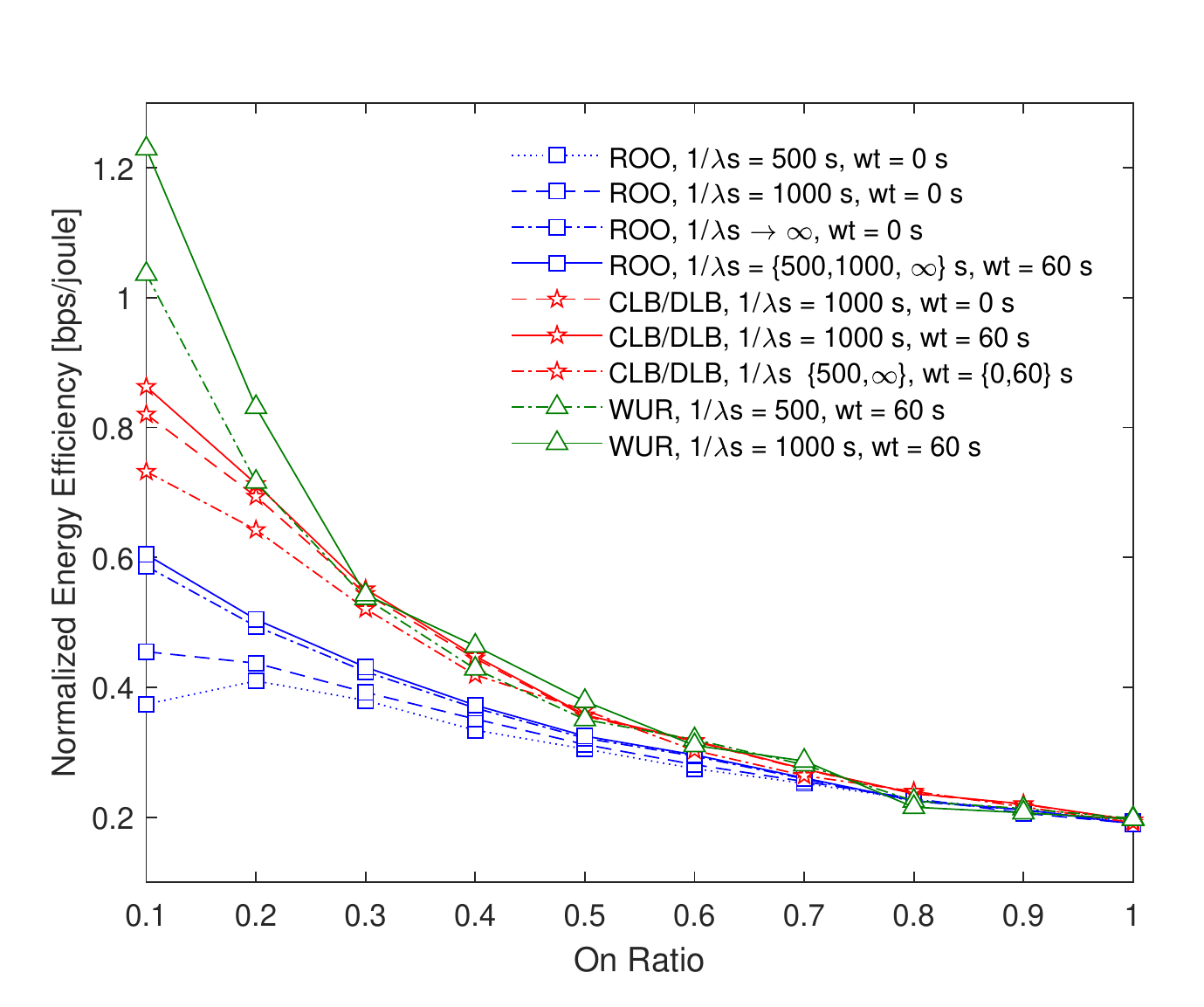}
\label{fig:lowutil_energy}}		
\caption{Average throughput and normalized energy efficiency along with on ratio (i.e., fraction of non-sleeping SBSs) for $1/\lambda_{\rm S}\,{\in}\,\{500\,\text{s}, 1000\,\text{s}, \infty\}$ and $w_{\rm t}\,{\in}\,\{0,60 \, \text{s}\}$ assuming low network utilization of $1\%$ (i.e., $1/\lambda_{\rm U} \,{=}\, 1000\,\text{s}$ and $1/\lambda_{\rm F} \,{=}\, 1\,\text{MB}$).}
\label{fig:lowutil_rate_energy}
\end{figure}

In Fig.~\ref{fig:lowutil_rate_energy}, we present the respective network throughput and normalized energy efficiency results. We observe that the network throughput performance in Fig.~\ref{fig:lowutil_rate_energy}\subref{fig:lowutil_rate} shows closely related behavior to the blocking probability results (i.e., network throughput increases with decreasing blocking probability, and vice versa). In particular, we observe no significant average throughput loss when as many as $40\%$ of SBSs are in sleeping states. On the other hand, the average throughput of ROO keeps decreasing continuously as more SBSs are put into sleeping states, which finally reads as high as $20\%$ throughput loss for non-sleeping SBSs fraction of $40\%$.

The normalized energy efficiency results in Fig.~\ref{fig:lowutil_rate_energy}\subref{fig:lowutil_energy} involve some interesting conclusions as follows. 1) Energy efficiency of ROO is worse than that of CLB/DLB whereas CLB/DLB is as energy-efficient as the more complex WUC scheme for non-sleeping SBS fractions greater than $30\%$. 2) Although ROO attains the maximum throughput only under nonzero tolerable delay (see Fig.~\ref{fig:lowutil_rate_energy}\subref{fig:lowutil_rate}), the maximum energy efficiency can be achieved under both zero and nonzero tolerable delays. In particular, while the maximum energy efficiency of ROO is invariant to sleep period under nonzero tolerable delay, the best sleep period turns out to be $\lambda_{\rm S}\,{\rightarrow}\,\infty$ under zero tolerable delay. As a result, \textit{the energy efficiency for ROO under zero tolerable delay gets maximized when OOS scheme is not applied dynamically} (i.e., no on/off events after initialization). 3) Although network throughput for CLB/DLB is maximized for $1/\lambda_{\rm S}\,{\in}\,\{500\,\text{s}, 1000\,\text{s}\}$ with a significant gap between the no dynamic OOS case (i.e., $\lambda_{\rm S}\,{\rightarrow}\,\infty$), the energy efficiency gets maximized only for $1/\lambda_{\rm S}\,{=}\,1000\,\text{s}$ under any choice of tolerable delay. \textit{Regardless of the particular tolerable delay in CLB/DLB, assigning short sleep time is therefore as energy inefficient as keeping SBSs in sleep states for very long}, which identifies an optimal sleep period in between.

\subsection{High Utilization Performance}

We now consider a high network utilization scenario with the UE service request rate of $1/\lambda_{\rm U} \,{=}\, 100\,\text{s}$ and the average file size of $1/\lambda_{\rm F} \,{=}\, 1\,\text{MB}$. The respective utilization is on the order of $20\%$.
We assume a representative finite sleep time period together with no dynamic OOS case, i.e., $1/\lambda_{\rm S}\,{\in}\,\{1000\,\text{s}, \infty\}$, together with both zero and nonzero tolerable delays, i.e., $w_{\rm t}\,{\in}\,\{0,60\,\text{s}\}$. In Fig.~\ref{fig:lowutil_pblock}, we present blocking probability results along with on ratio. As before, we observe that the performances of CLB and DLB are much better than that of ROO, and are the same as that of WUC whenever at least $50\%$ of the SBSs are non-sleeping. In addition, DLB has a close performance to CLB, as before. We also observe that the performance of any OOS algorithm improves together with either nonzero tolerable delay, or applying dynamic OOS (i.e., $1/\lambda_{\rm S}\,{=}\,1000\,\text{s}$ instead of $1/\lambda_{\rm S}\,{\rightarrow}\,\infty$) on top of that. Regardless of the particular OOS strategy, the blocking probabilities are observed to be higher than those in Fig.~\ref{fig:lowutil_pblock} as the fraction of non-sleeping SBSs decreases, which is basically due to the increased network utilization.           


\begin{figure}[!t]
	\centering
    \includegraphics[width=0.7\textwidth]{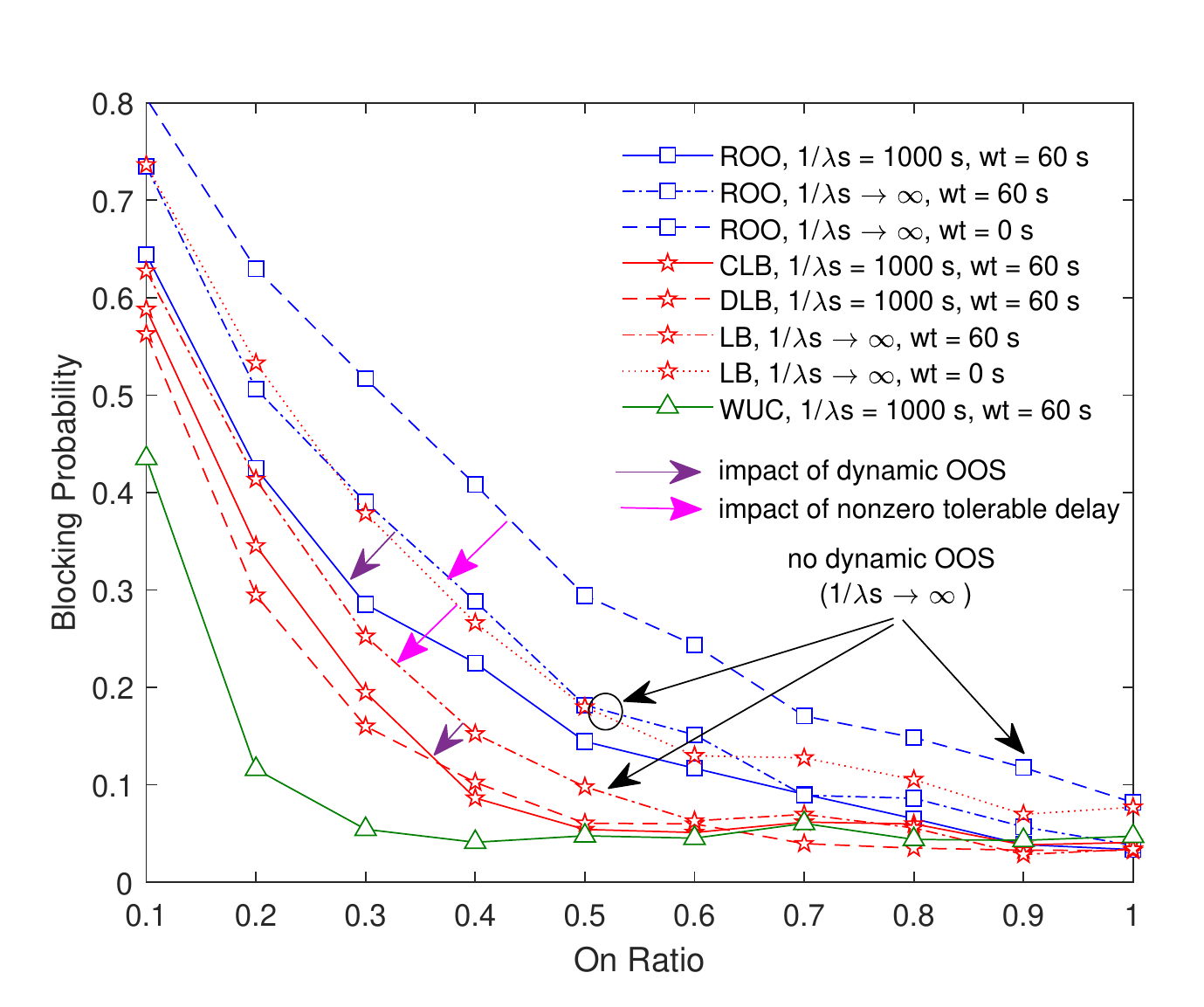}
	\caption{Blocking probability $\mathsf{P_{block}}$ along with on ratio (i.e., fraction of non-sleeping SBSs) for $1/\lambda_{\rm S}\,{\in}\,\{1000\,\text{s}, \infty\}$ and $w_{\rm t}\,{\in}\,\{0,60 \, \text{s}\}$ assuming low network utilization of $20\%$ (i.e., $1/\lambda_{\rm U} \,{=}\, 100\,\text{s}$ and $1/\lambda_{\rm F} \,{=}\, 2\,\text{MB}$).}
	\label{fig:highutil_pblock}
\end{figure}

\begin{figure}[!t]
\centering\hspace{-0.0in}
\subfloat[Average Throughput, $\mathsf{R_{SCN}}$] {\includegraphics[width=0.68\textwidth]{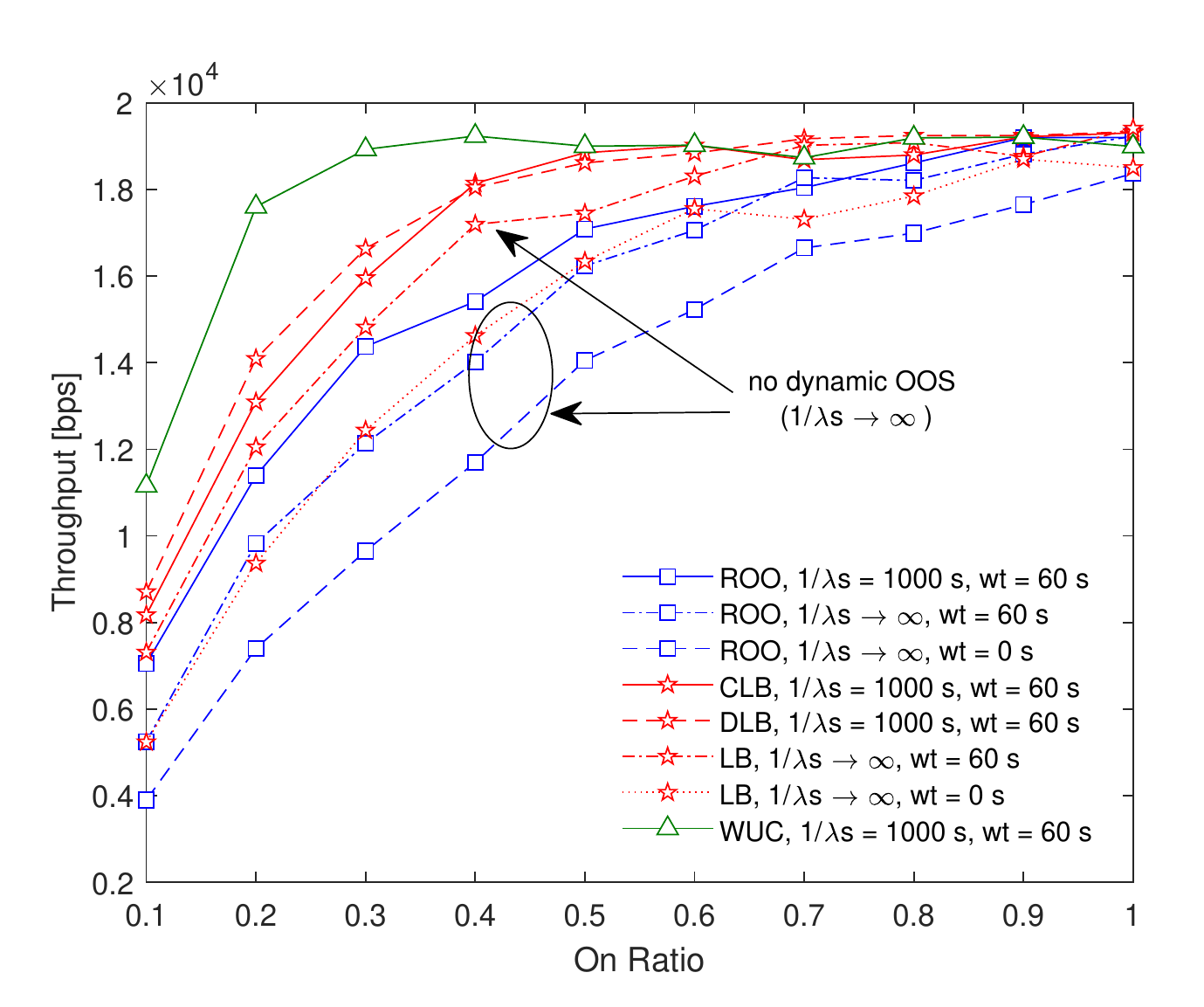}
\label{fig:highutil_rate}}\vspace{-0.3in}
\subfloat[Normalized Energy Efficiency, $\mathsf{EE}$] {\includegraphics[width=0.68\textwidth]{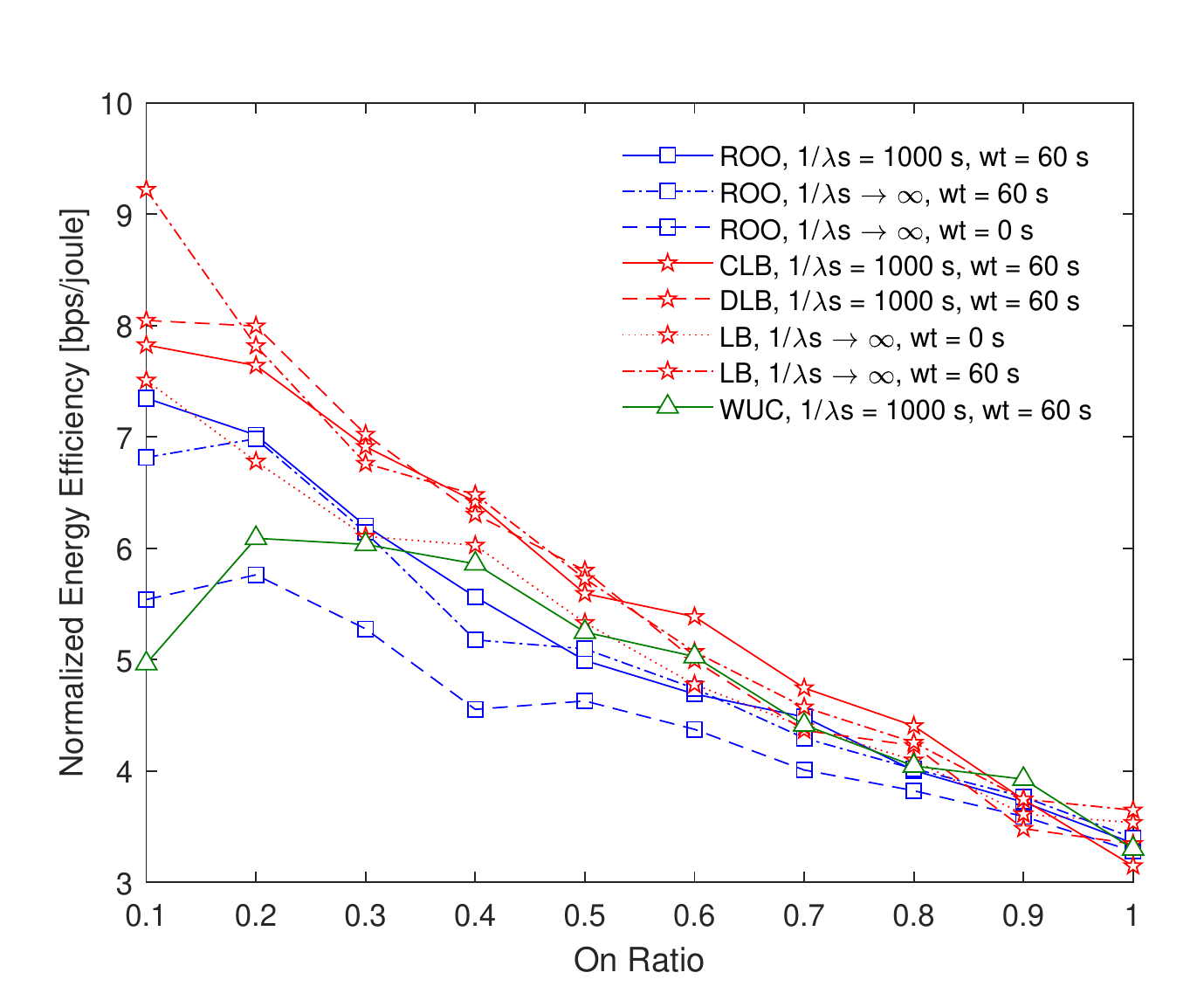}
\label{fig:highutil_energy}}		
\caption{Average throughput and normalized energy efficiency along with on ratio (i.e., fraction of non-sleeping SBSs) for $1/\lambda_{\rm S}\,{\in}\,\{1000\,\text{s}, \infty\}$
and $w_{\rm t}\,{\in}\,\{0,60 \, \text{s}\}$ assuming low network utilization of $20\%$ (i.e., $1/\lambda_{\rm U} \,{=}\, 100\,\text{s}$ and $1/\lambda_{\rm F} \,{=}\, 2\,\text{MB}$).}
\label{fig:highutil_rate_energy}
\end{figure}

In Fig.~\ref{fig:highutil_rate_energy}, we demonstrate the average throughput and normalized energy efficiency performances against on ratio. As before, the average throughput results in Fig.~\ref{fig:highutil_rate_energy}\subref{fig:highutil_rate} indicate that the performance of CLB and DLB are much better than that of ROO, and are the same as WUC for a broad range of non-sleeping SBS fractions (i.e., greater than $0.5$). In particular, the average throughput of either CLB or DLB remains almost unchanged even when $50\%$ of the SBSs are put into sleeping states, while the respective loss in ROO throughput appears to be between $10\%$-$30\%$ for the same on ratio. Note that the average throughput results in Fig.~\ref{fig:highutil_rate_energy}\subref{fig:highutil_rate} are much higher as compared to that of Fig.~\ref{fig:lowutil_rate_energy}\subref{fig:lowutil_rate} owing to the increased network utilization. In addition, the average throughput increases for all the OOS algorithms as UEs become more delay tolerant.  

We also present the respective normalized energy efficiency results in Fig.~\ref{fig:highutil_rate_energy}\subref{fig:highutil_energy} for this high utilization scenario. We observe that the energy efficiency of CLB and DLB gets maximized with the nonzero tolerable delay (i.e., $w_{\rm t}\,{\in}\, 60 \, \text{s}$), which is superior to not only ROO but also more sophisticated WUC scheme. This interesting result indicates that although the average network throughput is maximized (through decreasing blocking probabilities) by the deliberate wake-up control mechanism of WUC, the resulting scheme becomes less energy-efficient. In other words, while the network rejects less number of UE requests by further incorporating the sleeping SBSs, the overall network starts consuming more power since not all SBSs are allowed to complete their full sleep period. As a result, the energy efficiency of WUC deteriorates, and falls even below ROO under certain settings. We therefore conclude that, \textit{in contrast to low utilization scenario, the energy efficiency of WUC can be poor under high network utilization, although the associated average throughput might still be the best.}   


\section{Conclusion}\label{sec:conclusion}

In this study, we consider OOS strategies to have energy-efficient SCNs. In particular, we propose a novel load definition for the SCN traffic, and derived its approximate distribution rigorously. Two novel load based OOS algorithms (i.e., CLB and DLB) are also proposed together with two benchmark strategies ROO (i.e., simple baseline) and WUC (i.e., sophisticated). We show that CLB and DLB perform better than ROO, and have similar performance as compared to WUC under low traffic periods. Assuming high network utilization, CLB and DLB turns out to be even more energy-efficient then WUC. We finally show that the performance of CLB can be efficiently attained by DLB in a distributed fashion relying on the statistics of the traffic load. As a future work,  traffic load model can be extended to capture diverse mobile usage patterns by including the distributions of inter-arrival time, and file size distributions. Besides, wake-up control and load based schemes can be extended by considering mobile power consumption, and macrocell serving capacity in mid-traffic and high-traffic profiles.   


\bibliography{IEEEabrv,References}

\begin{thebibliography}{10}
\providecommand{\url}[1]{#1}
\csname url@samestyle\endcsname
\providecommand{\newblock}{\relax}
\providecommand{\bibinfo}[2]{#2}
\providecommand{\BIBentrySTDinterwordspacing}{\spaceskip=0pt\relax}
\providecommand{\BIBentryALTinterwordstretchfactor}{4}
\providecommand{\BIBentryALTinterwordspacing}{\spaceskip=\fontdimen2\font plus
\BIBentryALTinterwordstretchfactor\fontdimen3\font minus
  \fontdimen4\font\relax}
\providecommand{\BIBforeignlanguage}[2]{{%
\expandafter\ifx\csname l@#1\endcsname\relax
\typeout{** WARNING: IEEEtran.bst: No hyphenation pattern has been}%
\typeout{** loaded for the language `#1'. Using the pattern for}%
\typeout{** the default language instead.}%
\else
\language=\csname l@#1\endcsname
\fi
#2}}
\providecommand{\BIBdecl}{\relax}
\BIBdecl

\bibitem{bhushan2014network}
N.~Bhushan, J.~Li, D.~Malladi, R.~Gilmore, D.~Brenner, A.~Damnjanovic, R.~T.
  Sukhavasi, C.~Patel, and S.~Geirhofer, ``Network densification: the dominant
  theme for wireless evolution into {5G},'' \emph{IEEE Commun. Mag.}, vol.~52,
  no.~2, pp. 82--89, 2014.

\bibitem{quek2013small}
T.~Q. Quek, G.~de~la Roche, {\.I}.~G{\"u}ven{\c{c}}, and M.~Kountouris,
  \emph{Small cell networks: Deployment, {PHY} techniques, and resource
  management}.\hskip 1em plus 0.5em minus 0.4em\relax Cambridge University
  Press, 2013.

\bibitem{hwang2013holistic}
I.~Hwang, B.~Song, and S.~S. Soliman, ``A holistic view on hyper-dense
  heterogeneous and small cell networks,'' \emph{IEEE Commun. Mag.}, vol.~51,
  no.~6, pp. 20--27, 2013.

\bibitem{samarakoon2016ultra}
S.~Samarakoon, M.~Bennis, W.~Saad, M.~Debbah, and M.~Latva-aho, ``Ultra dense
  small cell networks: Turning density into energy efficiency,'' \emph{IEEE J.
  Select. Areas Commun. (JSAC)}, vol.~34, no.~5, pp. 1267--1280, 2016.

\bibitem{li2014throughput}
C.~Li, J.~Zhang, and K.~B. Letaief, ``Throughput and energy efficiency analysis
  of small cell networks with multi-antenna base stations,'' \emph{IEEE Trans.
  Wireless Commun.}, vol.~13, no.~5, pp. 2505--2517, 2014.

\bibitem{soh2013energy}
Y.~S. Soh, T.~Q. Quek, M.~Kountouris, and H.~Shin, ``Energy efficient
  heterogeneous cellular networks,'' \emph{IEEE J. Select. Areas Commun.
  (JSAC)}, vol.~31, no.~5, pp. 840--850, 2013.

\bibitem{merwaday2016optimisation}
A.~Merwaday and I.~G{\"u}ven{\c{c}}, ``Optimisation of {FeICIC} for energy
  efficiency and spectrum efficiency in {LTE}-advanced {HetNets},'' \emph{IET
  Electronics Letters}, vol.~52, no.~11, pp. 982--984, 2016.

\bibitem{GreenBSN:EnergyPropCell}
C.~Peng, S.-B. Lee, S.~Lu, and H.~Luo, ``Green{BSN}: Enabling
  energy-proportional cellular base station networks,'' \emph{IEEE Trans.
  Mobile Computing}, vol.~13, no.~11, pp. 2537--2551, Nov 2014.

\bibitem{5GAccess:linkProvisioning}
P.~Leroux, A.~Callard, N.-D. Dao, and A.~Stephenne, ``{5G} access-link
  provisioning and coordination: Tradeoff between proactive and reactive
  strategies,'' in \emph{Proc. IEEE Wireless Commun. Netw. Conf. (WCNC)}, Mar.
  2015, pp. 995--999.

\bibitem{zeng2013directional}
Y.~Zeng, K.~Xiang, D.~Li, and A.~V. Vasilakos, ``Directional routing and
  scheduling for green vehicular delay tolerant networks,'' \emph{Wireless
  Networks}, vol.~19, no.~2, pp. 161--173, 2013.

\bibitem{he2013emd}
S.~He, X.~Li, J.~Chen, P.~Cheng, Y.~Sun, and D.~Simplot-Ryl, ``{EMD}:
  energy-efficient {P2P} message dissemination in delay-tolerant wireless
  sensor and actor networks,'' \emph{IEEE J. Select. Areas Commun. (JSAC)},
  vol.~31, no.~9, pp. 75--84, 2013.

\bibitem{cao2013routing}
Y.~Cao and Z.~Sun, ``Routing in delay/disruption tolerant networks: A taxonomy,
  survey and challenges,'' \emph{IEEE Commun. Surveys \& Tutorials}, vol.~15,
  no.~2, pp. 654--677, 2013.

\bibitem{loadBasedOnOffHaluk}
H.~Celebi and I.~Guvenc, ``Load analysis and sleep mode optimization for
  energy-efficient {5G} small cell networks,'' in \emph{Proc. IEEE Int. Conf.
  Commun. (ICC) Workshops}, May 2017, pp. 1159--1164.

\bibitem{Natarajan2016SmaCel}
C.~Liu, B.~Natarajan, and H.~Xia, ``Small cell base station sleep strategies
  for energy efficiency,'' \emph{IEEE Trans. Vehic. Technol.}, vol.~65, no.~3,
  pp. 1652--1661, Mar. 2016.

\bibitem{Ashraf2011SleMod}
I.~Ashraf, F.~Boccardi, and L.~Ho, ``{SLEEP} mode techniques for small cell
  deployments,'' \emph{IEEE Commun. Mag.}, vol.~49, no.~8, pp. 72--79, Aug.
  2011.

\bibitem{Vereecken2012TheEff}
W.~Vereecken, I.~Haratcherev, M.~Deruyck, W.~Joseph, M.~Pickavet, L.~Martens,
  and P.~Demeester, ``The effect of variable wake up time on the utilization of
  sleep modes in femtocell mobile access networks,'' in \emph{Proc. Annual
  Conf. Wireless On-Demand Network Systems and Services (WONS)}, Jan. 2012, pp.
  63--66.

\bibitem{tanemura2003statistical}
M.~Tanemura, ``Statistical distributions of poisson voronoi cells in two and
  three dimensions,'' \emph{FORMA}, vol.~18, no.~4, pp. 221--247, 2003.

\bibitem{ferenc2007size}
J.-S. Ferenc and Z.~N{\'e}da, ``On the size distribution of poisson voronoi
  cells,'' \emph{Physica A: Statistical Mechanics and its Applications}, vol.
  385, no.~2, pp. 518--526, 2007.

\bibitem{davies1974size}
C.~Davies, ``Size distribution of atmospheric particles,'' \emph{Journal of
  Aerosol Science}, vol.~5, no.~3, pp. 293--300, 1974.

\bibitem{Ross2009IntProb}
S.~M. Ross, \emph{Introduction to Probability Models}, 10th~ed.\hskip 1em plus
  0.5em minus 0.4em\relax Academic Press, 2009.

\bibitem{milios2009probability}
D.~Milios, ``Probability distributions as program variables,'' \emph{Master's
  thesis, School of Informatics, University of Edinburgh}, 2009.

\bibitem{fewell2006area}
M.~Fewell, ``Area of common overlap of three circles,'' Defence Science and
  Technology Organisation, Report DSTO-TN-0722, Oct. 2006.

\bibitem{blogowski2012dimensioning}
A.~Blogowski, O.~Klopfenstein, and B.~Renard, ``Dimensioning {X2} backhaul link
  in {LTE} networks,'' in \emph{Proc. IEEE Int. Conf. Commun. (ICC)}, 2012, pp.
  2768--2773.

\end{thebibliography}
\bibliographystyle{IEEEtran}

\end{document}